\documentclass[useAMS,usenatbib]{mn2e}
\usepackage{mathrsfs,amsmath,graphics,color,epsfig,amssymb}
\usepackage{mathptmx}
\usepackage{epsfig}
\usepackage{booktabs}
\usepackage{rotating,rotfloat}
\usepackage{longtable}
\usepackage{graphicx}
\usepackage{url} 
\usepackage{listings}
\usepackage{color, soul}
\usepackage{pslatex} 
\usepackage{booktabs}
\usepackage{dcolumn}
\usepackage{multicol}
\usepackage{subcaption}
\usepackage{multirow}
\usepackage[hyperindex,breaklinks=true, colorlinks, citecolor=blue]{hyperref}
\usepackage{morefloats}
\usepackage{bm}

\newcommand{\wmap}{WMAP\:}
\newcommand{\planck}{\textit{Planck}\:}
\newcommand{\iras}{IRAS}

\newcommand{\commander}{\texttt{COMMANDER}\:}

\def\ba{\begin{eqnarray}}
\def\ea{\end{eqnarray}}
\def\be{\begin{equation}}
\def\ee{\end{equation}}

\def\bdelta{{\delta }}

\makeatletter
\def\ref@jnl#1{{\rmfamily#1}}%
\newcommand\aj{\ref@jnl{AJ}}%
\newcommand\araa{\ref@jnl{ARA\&A}}%
\newcommand\apj{\ref@jnl{ApJ}}%
\newcommand\apjl{\ref@jnl{ApJ}}%
\newcommand\apjs{\ref@jnl{ApJS}}%
\newcommand\apss{\ref@jnl{Ap\&SS}}%
\newcommand\aap{\ref@jnl{A\&A}}%
\newcommand\aapr{\ref@jnl{A\&A~Rev.}}%
\newcommand\aaps{\ref@jnl{A\&AS}}%
\newcommand\baas{\ref@jnl{BAAS}}%
\newcommand\memras{\ref@jnl{MmRAS}}%
\newcommand\mnras{\ref@jnl{MNRAS}}%
\newcommand\pra{\ref@jnl{Phys.~Rev.~A}}%
\newcommand\prb{\ref@jnl{Phys.~Rev.~B}}%
\newcommand\prc{\ref@jnl{Phys.~Rev.~C}}%
\newcommand\prd{\ref@jnl{Phys.~Rev.~D}}%
\newcommand\pre{\ref@jnl{Phys.~Rev.~E}}%
\newcommand\prl{\ref@jnl{Phys.~Rev.~Lett.}}%
\newcommand\pasp{\ref@jnl{PASP}}%
\newcommand\pasj{\ref@jnl{PASJ}}%
\newcommand\ssr{\ref@jnl{Space~Sci.~Rev.}}%
\newcommand\nat{\ref@jnl{Nature}}%
\newcommand\iaucirc{\ref@jnl{IAU~Circ.}}%
\newcommand\aplett{\ref@jnl{Astrophys.~Lett.}}%
\newcommand\apspr{\ref@jnl{Astrophys.~Space~Phys.~Res.}}%
\newcommand\nphysa{\ref@jnl{Nucl.~Phys.~A}}%
\newcommand\physrep{\ref@jnl{Phys.~Rep.}}%
\newcommand\planss{\ref@jnl{Planet.~Space~Sci.}}%
\newcommand\procspie{\ref@jnl{Proc.~SPIE}}%
\newcommand\pasa{\ref@jnl{Pub.~Ast.~Soc.~Aus.}}

\def\pdeg{\ifmmode $\setbox0=\hbox{$^{\circ}$}\rlap{\hskip.11\wd0 .}$^{\circ}
          \else \setbox0=\hbox{$^{\circ}$}\rlap{\hskip.11\wd0 .}$^{\circ}$\fi}
\providecommand{\sorthelp}[1]{}

\makeatletter

\makeatletter
\newcommand\footnoteref[1]{\protected@xdef\@thefnmark{\ref{#1}}\@footnotemark}
\makeatother

\title[\mbox{C-BASS} Diffuse Microwave Foregrounds]{The C-Band All-Sky Survey (\mbox{C-BASS}): Template Fitting of Diffuse Galactic Microwave Emission in the Northern Sky}

\author[S.\,E.~Harper et al.]
  {S.\,E.~Harper,$\!^1$\thanks{E-mail: stuart.harper@manchester.ac.uk}
    C.~Dickinson,$\!^{1,2}$ 
    A.~Barr,$\!^1$ 
    R.~Cepeda-Arroita,$\!^1$ 
    R.\,D.\,P.~Grumitt,$\!^{3}$ \newauthor
    H.\,M.~Heilgendorff,$^{6}$
    L.~Jew,$\!^{3}$
    J.\,L.~Jonas,$\!^{7,8}$
    M.\,E.~Jones,$\!^3$
    J.\,P.~Leahy,$\!^1$
    J.~Leech,$\!^{3}$\newauthor
    T.\,J.~Pearson,$\!^{2}$
    M.\,W.~Peel,$\!^{4,5}$
    A.\,C.\,S.~Readhead,$\!^{2}$
    A.\,C.~Taylor$\!^3$\newauthor 
  \\
$^1$Jodrell Bank Centre for Astrophysics, Alan Turing Building, Department of Physics \& Astronomy, School of Natural Sciences, \\ The University of Manchester,
Oxford Road, Manchester, M13 9PL, U.K. \\ 
$^{2}$Cahill Centre for Astronomy and Astrophysics, California Institute of Technology, Pasadena, CA 91125, U.S.A. \\
$^3$Sub-department of Astrophysics, University of Oxford, Denys Wilkinson Building, Keble Road, Oxford OX1 3RH, U.K. \\
$^{4}$Instituto de Astrof\'{i}sica de Canarias, E-38205 La Laguna, Tenerife, Spain\\
$^{5}$Departamento de Astrof\'{i}sica, Universidad de La Laguna (ULL), E-38206 La Laguna, Tenerife, Spain\\
$^{6}$School of Mathematics, Statistics \& Computer Science, University of KwaZulu-Natal, Westville Campus, Private Bag X54001,\\ Durban, South Africa \\
$^{7}$Department of Physics and Electronics, Rhodes University, Grahamstown, 6139, South Africa \\
$^{8}$South African Radio Astronomy Observatory, 2 Fir Road, Observatory, Cape Town, 7925, South Africa\\
}

\begin{document}


\setlength{\topmargin}{-15mm}

\pagerange{\pageref{firstpage}--\pageref{lastpage}} \pubyear{2002}

\maketitle

\label{firstpage}

\begin{abstract}

The C-Band All-Sky Survey (\mbox{C-BASS}) has observed the Galaxy at 4.76\,GHz with an angular resolution of $0\pdeg73$ full-width half-maximum, and detected Galactic synchrotron emission with high signal-to-noise ratio over the entire northern sky ($\delta > -15^{\circ}$). We present the results of a spatial correlation analysis of Galactic foregrounds at mid-to-high ($b > 10^\circ$) Galactic latitudes using a preliminary version of the C-BASS  intensity map. We jointly fit for synchrotron, dust, and free-free components between $20$ and $1000$\,GHz and look for differences in the Galactic synchrotron spectrum, and the emissivity of anomalous microwave emission (AME) when using either the \mbox{C-BASS} map or the 408\,MHz all-sky map to trace synchrotron emission. We find marginal evidence for a steepening ($\left<\Delta\beta\right> = -0.06\pm0.02$) of the Galactic synchrotron spectrum at high frequencies resulting in a mean spectral index of $\left<\beta\right> = -3.10\pm0.02$ over $4.76$--$22.8$\,GHz.  Further, we find that the synchrotron emission can be well modelled by a single power-law up to  a few tens of GHz. Due to this, we find that the AME emissivity is not sensitive to changing the synchrotron tracer from the 408\,MHz map to the 4.76\,GHz map. We interpret this as strong evidence for the origin of AME being spinning dust emission.

\end{abstract}

\section{Introduction}\label{sec:intro}

Large-scale diffuse Galactic radio emission, at frequencies \mbox{$0.1$--$50$\,GHz}, is composed of three principal components: synchrotron emission from the propagation of cosmic rays through the Galactic magnetic field \citep[e.g.,][]{Strong2011}, thermal free-free emission from ionized gas \citep[e.g.,][]{Dickinson2003}, and spinning dust emission from the rapid rotation of small grains in the interstellar medium (ISM) \citep[e.g.,][]{Dickinson2018}. The study of these Galactic components is important both for understanding the astrophysics of our Galaxy, and also for studies of the cosmic microwave background (CMB) \citep[e.g.,][]{Oliveira-Costa2008,Bonaldi2011,Remazeilles2016}.

There are only a limited number of radio surveys that preserve the large-scale structure of Galactic emission. The all-sky 408\,MHz map \citep{Haslam1982} has become the standard map for tracing Galactic synchrotron emission as it has  little impact from both synchrotron self-absorption and free-free contamination. However, the 408\,MHz map is limited by being total intensity only and having numerous systematics that are challenging to quantify \citep{Remazeilles2015}. Other radio surveys of the large-scale Galactic structure at 820\,MHz \citep{Berkhuijsen1972}, 1420\,MHz \citep{Reich2001,Calabretta2014}, and 2.3\,GHz \citep{Reif1987,Jonas1998} are also limited due either to partial sky coverage, systematics, and being in total intensity only ($I+Q$ for the 2.3\,GHz map). The S-band Polarization All-Sky Survey \citep[S-PASS;][]{Carretti2019} provided the first map of polarized emission in the Southern sky using the Parkes telescope at 2.3\,GHz. New data from the Q-U-I JOint Tenerife (QUIJOTE) experiment at $10$--$20$\,GHz will complement these surveys in the northern hemisphere \citep{Genova2015,Guidi2020}.

The limited number of surveys at the intermediate frequencies between 408\,MHz and the lowest frequency Wilkinson Microwave Anisotropy Probe (WMAP) 22.8\,GHz band means that capturing spectral curvature in the synchrotron component is almost impossible. We expect to see spectral steepening in the synchrotron emission due to energy losses in the cosmic ray electron (CRE) population \citep[e.g.,][]{Strong2011} or potentially even flattening due to re-energisation of the CRE population from active star formation regions \citep{Bennett2003}. The ARCADE2 results suggest that there is evidence for synchrotron spectral curvature \citep{Kogut2011}, however the ARCADE2 survey was limited to a relatively small portion of the sky.

There are also still many unknown quantities regarding dust emission at radio frequencies, commonly referred to as either anomalous microwave emission (AME) \citep{Kogut1996b,Leitch1997} or, when linked with polycyclic aromatic hydrocarbons (PAHs) or very small grains, spinning dust emission \citep{Draine1998a,AliHaimoud2009}. The \textit{Planck} \commander analysis has provided the current best all-sky map of AME \citep{planck2013-p06,planck2016-l04}. However, it was found that the \commander analysis derived AME amplitude can differ  by 30--50\,per\,cent when compared to other analyses due to correlations with free-free and synchrotron emission components even in mid-to-high latitude ($b > 10^\circ$) regions \citep{planck2014-a31}. Similarly \citet{Cepeda-Arroita2021} found significant differences in the fitted AME amplitude when using both the \mbox{C-BASS} and 408\,MHz data to constrain low frequencies rather than just the 408\,MHz data alone in the $\lambda$-Orionis region.

The study of AME, free-free and synchrotron emission at mid-to-high latitudes can be greatly improved by using a cleaner estimate of the synchrotron emission at WMAP and \textit{Planck} frequencies. \mbox{C-BASS} offers just such as estimate as it is nearer in frequency, meaning that it will be less impacted by spectral curvature. Further advantages of the \mbox{C-BASS} map over the 408\,MHz map are a well-understood beam response, allowing the map to be deconvolved to a known Gaussian beam pattern, and well-understood noise and systematics; all of which are not available for the 408\,MHz map. At present only the northern part of the \mbox{C-BASS} survey is complete, but this will be extended to the full-sky in the future.

In this paper we use a preliminary version of the \mbox{C-BASS} intensity map of the northern sky to perform a pixel-space template fitting correlation analysis, a method that has a long heritage for studying Galactic foregrounds \citep[e.g.,][]{deOliveira-Costa1997,Bennett2003,Banday2003,Davies2006,Ghosh2012,Peel2012}. We use the \mbox{C-BASS} map in combination with the 408\,MHz map to estimate the degree of spectral curvature across the sky in regions of approximately 100\,sq.\,degrees. We also investigate if there is any change in the fitted dust coefficients at 22.8\,GHz when using \mbox{C-BASS} in place of the 408\,MHz map, and look for any evidence of a hard synchrotron component that may explain some or all of the observed AME at mid-to-high latitudes \citep{Peel2012}. Finally, we look to see if there is any preference for any particular dust tracer when measuring AME at mid-to-high latitudes.

The paper is organized as follows: In \autoref{sec:cbass} we provide an overview of the \mbox{C-BASS} experiment, data reduction methods, and map-making. In \autoref{sec:data} we describe all the ancillary datasets used in this analysis. In \autoref{sec:templatefitting} we outline the template fitting method as well as the masks and regions that we use to divide up the sky. In \autoref{sec:results} we outline the results of the template fitting analysis. In \autoref{sec:fitting} we fit model spectra of spinning dust and thermal dust to the dust template fitting coefficients. Finally, in \autoref{sec:discussion} we discuss the results of the template fitting analysis in a wider context before giving concluding remarks and a summary in \autoref{sec:conclusion}.

\vspace{-0.1cm}
\section{C-BASS North}\label{sec:cbass}

In this work we will use a preliminary version of the \mbox{C-BASS} intensity data. The \mbox{C-BASS} map presented here is identical to that used in \cite{Dickinson2019} and \cite{Cepeda-Arroita2021} except for a global calibration correction resulting in a 2.1\,per\,cent decrease in brightness; there is also a decrease in calibration uncertainty from 5 to 3\,per\,cent. This map will be almost identical to the final \mbox{C-BASS} intensity map except for small changes in data selection and processing aimed at eliminating subtle artefacts in the polarization data, which have negligible effect on the intensity maps. In the following sections we give an overview of the \mbox{C-BASS} data reduction pipeline. More details will be given in the upcoming \mbox{C-BASS} survey papers \mbox{(Taylor et al., in prep.)} and \mbox{(Pearson et al., in prep.)}.

\subsection{Northern Survey and Instrument} 

 The C-BASS project aims to map the entire sky in intensity and polarization at 4.76\,GHz \citep{Jones2018}. The northern survey used a 6.1\,m dish based at the Owens Valley Radio Observatory, with a nominal full-width half-maximum (FWHM) resolution of $0\pdeg73$. The antenna uses a Gregorian optical configuration that was designed to minimize sidelobe power by both under-illuminating the primary reflector and surrounding it in a radio absorbing baffle. Furthermore, to ensure a circularly symmetric beam pattern, the support struts for the secondary reflector were removed and it was instead supported by a low-loss dielectric foam cone \citep{Holler2013}. 

The \mbox{C-BASS} instrument is a dual circularly polarized correlation radiometer  that can obtain instantaneous measurements of Stokes parameters I, Q, and U. The receiver's nominal bandpass is $4.5$--$5.5$\,GHz but notch filters were used to suppress local radio frequency interference (RFI), reducing the effective bandwidth to 0.5\,GHz. 
Receiver stability is maintained by a continuous comparison between the sky and a resistive load that minimizes $1/f$ noise fluctuations. For details of the C-BASS receiver system see \citet{King2014}. 

The \mbox{C-BASS} northern survey observations were taken between 2012 and 2015. The survey scanning strategy slewed the telescope at a set of fixed elevations, between $37\pdeg2$ and $77\pdeg2$, over full $360^{\circ}$ sweeps in azimuth. The slewing rate was $\approx 4^{\circ}$\,s$^{-1}$ but was varied slightly between observations to average out any possible scan-synchronous instrumental systematics from the final \mbox{C-BASS} map. This observing strategy maps all declinations above $-15^{\circ}$, and covers approximately a 26000\,deg$^2$ area of sky. A full description of the northern survey maps will be given in \mbox{Taylor et al. (in prep.)}.

\subsection{\mbox{C-BASS} Data Reduction}

The \mbox{C-BASS} data are processed through a standard pipeline procedure, the details of which will be outlined in forthcoming papers \mbox{(Pearson et al., Taylor et al, in prep.)}. However, we provide a brief summary here. The key tasks of the pipeline are to flag sources of RFI and solar system objects, calibrate the data to brightness temperature units, and remove a number of systematic effects from the data. The two main systematics that are modelled and tracked are: a microphonics signal caused by the cryocooler system, which induces a 1.2\,Hz signal into the time-ordered data (TOD); and emission from the ground detected in the far sidelobes. 

Relative calibration is performed by injecting a regular noise diode signal into the front-end of the receiver. The noise diode is found to be stable to the level of 1\,per\,cent or better over many months, resulting in a relative calibration better than 1\,per\,cent in total. The calibration to the astronomical brightness scale is done using daily observations of \mbox{Tau\,A} and \mbox{Cas\,A} using the WMAP derived models for the flux density of these sources \citep{Weiland2011}. The bandpass-weighted central frequency of the \mbox{C-BASS} North intensity data is 4.76\,GHz when calibrated to a flat-spectrum ($\beta = -2$\footnote{Defined in brightness temperature units as $T_b \propto \nu^\beta$.}) source. Over the realistic range of source spectral indices observed in the \mbox{C-BASS} map ($-3.5 < \beta < -2$), colour corrections will contribute less than a 1\,per\,cent uncertainty. One remaining source of uncertainty is the primary beam deconvolution; tests of sources flux densities in the final \mbox{C-BASS} map suggest that the deconvolution (\autoref{sec:deconvolution}) has an additional 1\,per\,cent uncertainty. However, at present, the \mbox{C-BASS} calibration is still being finalized in preparation for the public release of the survey \mbox{(Taylor et al., in prep.)}, as such we adopt a conservative 3\,per\,cent calibration uncertainty for this work.

\subsection{\mbox{C-BASS} maps}

\subsubsection{Map Making}

To produce the \mbox{C-BASS} maps we use the destriping map-maker \texttt{Descart}~\citep{Sutton2010}. We use a destriping offset length of 5\,seconds to remove any large-scale $1/f$ noise in the data. The \mbox{C-BASS} map does not include  day-time data due to the impact of the Sun in the far-sidelobes of the beam on the large scale structures in the map. The final map has a $0\pdeg73$ FWHM, with a sensitivity of approximately 0.25\,mK/beam (instrumental white noise only). To estimate the level of the residual $1/f$ noise in the map we performed a jack-knife test where we split the \mbox{C-BASS} observations into two approximately equal sized datasets, and produced a map for each. Differencing the two maps, we compared the ratio of the root mean squared (RMS) estimated in the residual data (which contains both residual $1/f$ and white noise) to the expected RMS assuming just white noise. We find that on average the \mbox{C-BASS} map has a 10\,per\,cent excess of residual $1/f$ noise at scales of a few degrees and larger, which is negligible relative to other sources of uncertainty.

\subsubsection{Deconvolution}\label{sec:deconvolution}

The \mbox{C-BASS} beam is diffraction-limited with a main beam efficiency of 72.8\,per\,cent (i.e., the power within the first null, which is at $1\pdeg0$). The sidelobe structures of the \mbox{C-BASS} beam are imprinted into the final map and result in an effective calibration which varies with angular scale. In previous \mbox{C-BASS} papers \citep{Irfan2015,Dickinson2019,Cepeda-Arroita2021} this resulted in an effective calibration uncertainty due to the beam of approximately 5\,per\,cent.

We have accurately characterized the \mbox{C-BASS} beam using detailed physical optics simulations verified by observations of bright point sources and direct beam measurements using a radio transmitter, improving the measurements presented by \citet{Holler2013}. This allows us to deconvolve the effect of the beam, resulting in an effective Gaussian beam and a window function that is largely flat in log-harmonic space.
 
The deconvolution of the \mbox{C-BASS} map is done in spherical harmonic space using the routines provided by \texttt{HEALPix} \citep{gorski2005}. First, we transform the \mbox{C-BASS} beam model into a spherical harmonic transfer function ($B_\ell$). Next, we generate the transfer function for the $1\pdeg0$ Gaussian beam that we wish to smooth to ($G_\ell$) and divide this by the derived \mbox{C-BASS} beam transfer function. More details of the beam transfer function can be found in the \mbox{C-BASS} northern survey paper \mbox{(Taylor et al., in prep.)}. Finally, we multiply the ratio of the transfer function with the spherical harmonic amplitudes of the map such that the deconvolved \mbox{C-BASS} map is
\begin{equation}
    m(\theta,\phi) = \sum_{\ell=0}^{\infty}\sum_{m=-\ell}^{\ell} a_{\ell,m} \frac{G_\ell}{B_\ell} Y_\ell^m(\theta,\phi),
\end{equation}
where $m(\theta,\phi)$ is the map along the line-of-sight defined by $\theta$ and $\phi$, $G_\ell$ is a $1\pdeg0$ Gaussian beam transfer function, $B_\ell$ is the \mbox{C-BASS} beam transfer function, and $Y_\ell^m(\theta,\phi)$ are spherical harmonics. We find that after beam deconvolution the calibration of the map on all angular scales is 1\,per\,cent or better, with the remaining uncertainty due to small asymmetries in the beam. 

\subsubsection{Background Source Subtraction}\label{sec:source_sub}

After deconvolution of the map we subtract a model of the extragalactic background point sources. At the frequency and resolution of the \mbox{C-BASS} data the sky is confusion-limited at the level of $\approx 0.7$\,mK\,deg$^{-1}$ (\autoref{sec:noise} for details), which is several times greater than the instrumental noise level in the map. To remove the sources we use a combination of catalogues. We use the \mbox{C-BASS}-derived point source catalogue described in \citet{Grumitt2020} for all sources that are detected at 10\,$\sigma$ or better and do not lie within $|b| < 1^{\circ}$ of the Galactic plane. For fainter sources, or sources with a poor detection using \mbox{C-BASS} itself, we use several \mbox{C-band} point source catalogues: the Green Bank 6\,cm (GB6) survey \citep{Gregory1996}, the Parkes-MIT-NRAO (PMN) survey \citep{Wright1994} (for low declinations), and the \mbox{RATAN-800} survey \citep{Mingaliev2007} for sources missed by the GB6 survey around the North Celestial Pole (NCP). 

The sources' flux densities were binned in $a_{\ell m}$ space and multiplied by the beam transfer function as
\begin{equation}
    a_{\ell m} = \sum_{i=1}^{N_s} S_i Y_\ell^m(\theta_i,\phi_i) B_\ell,
\end{equation}
where $S_i$ is the source flux density from one of the catalogues, $N_s$ is the number of sources in the catalogue, $B_\ell$ is the transfer function for the deconvolved \mbox{C-BASS} map at $1^{\circ}$ resolution, and $Y_\ell^m$ are spherical harmonics. We note that for some sources, specifically those near the Galactic plane or with large flux densities, this method will not result in the perfect subtraction of the source  because of small pointing offsets, changes in source flux densities, or systematic errors in the measurement of the source flux density. In this analysis we mask  bright sources with $S_{4.76\mathrm{GHz}} > 1$\,Jy. \autoref{sec:mask} provides more details on masking.

\vspace{-0.1cm}
\section{Ancillary Data}\label{sec:data}

Here we provide an overview of all the ancillary datasets. All datasets are smoothed to a common resolution of $1^{\circ}$ FWHM, using beam transfer functions where available, i.e., for C-BASS, WMAP, and {\it Planck}. For WMAP and \textit{Planck} maps the CMB has been subtracted using the Spectral Matching Independent Component Analysis (SMICA) estimate from \citet{planck2013-p06}. A summary of all the datasets used is given in \autoref{tab:summary}.

\begin{table*}
\caption{Datasets used in this analysis. The first block of maps are used as templates for fitting the synchrotron, free-free, and dust components. The second block lists the maps to which we applied the template fitting procedure.}
\begin{tabular}{lcccll}
\hline\hline
Telescope/Survey   		&Frequency   	&FWHM       & $\sigma_\mathrm{cal}$    &Reference   	                    &Notes   	\\
                      & (GHz)       & (arcmin)  & (\%)            & & \\
\hline
Haslam			        	&0.408		         &51	                	&10          & \citet{Remazeilles2015}	&Synchrotron template \\
C-BASS			        	&4.76	            	&44		                &3	     & This work			& Synchrotron template \\
WHAM H$\alpha$		  	& N/A		           & 1	                 	& 10	     & \citet{Dickinson2003,Finkbeiner2003} 	&Free-free template \\
\planck  HFI 353\,GHz	&353		&4.7		&5	     &\citet{planck2016-l01}    &Dust template 	\\
\planck  353\,GHz optical depth	&353		&5.0		&5	     &\citet{planck2013-p06b}    &Dust template	\\
\planck dust radiance $\Re$		&--		&5.0		&--	&\citet{planck2013-p06b}  & Dust template \\
\iras 100\,$\mu$m 	&2997		&4.3		&13.5        &\citet{Miville-Deschenes2005}  	&Dust template \\
FDS8 model 8				&94		&6.1		&--		&\citet{Schlegel1998}			&Dust template		\\
\hline
\wmap K-band	&22.8		&51.3		&3	     &\citet{Bennett2013}    	&9-year 	\\
\planck  LFI 30\,GHz	&28.4		&33.16		&3	     &\citet{planck2016-l01}    &PR3	\\
\wmap Ka-band	&33.0		&39.1		&3           &\citet{Bennett2013}    	&9-year 	\\
\wmap Q-band	&40.7		&30.8		&3	     &\citet{Bennett2013}    	&9-year \\
\planck  LFI 44\,GHz	&44.1		&28.09		&3	     &\citet{planck2016-l01}    &PR3	\\
\wmap V-band	&60.7		&30.8		&3	     &\citet{Bennett2013}    	&9-year \\
\planck  LFI 70\,GHz	&70.4		&13.08		&3	&\citet{planck2016-l01}    &PR3	\\
\wmap W-band	&93.5		&30.8		&3	&\citet{Bennett2013}    	&9-year \\
\planck  HFI 143\,GHz	&143		&7.18		&5	&\citet{planck2016-l01}    &PR3	\\
\planck  HFI 217\,GHz	&217		&4.87		&5	&\citet{planck2016-l01}    &PR3	\\
\planck  HFI 353\,GHz	&353		&4.7		&5	&\citet{planck2016-l01}    &PR3	\\
\planck  HFI 545\,GHz	&545		&4.73		&5	&\citet{planck2016-l01}    &PR3	\\
\planck HFI 857\,GHz	&857		&4.51		&5	&\citet{planck2016-l01}    &PR3	\\
\iras 100\,$\mu$m 	  &2997		&4.3		&13.5        &\citet{Miville-Deschenes2005}  	& IRIS \\
\hline
\end{tabular}
\label{tab:summary}
\end{table*}

\subsection{Synchrotron Templates}

As well as using the \mbox{C-BASS} map (\autoref{sec:cbass}) to trace the diffuse synchrotron emission we also use the 408\,MHz radio continuum survey by \citet{Haslam1982}, which has remained for decades the best tracer of diffuse Galactic synchrotron emission at WMAP and \textit{Planck} frequencies. The survey was undertaken during the 1960s and 1970s and combines observations from the Effelsberg 100\,m, and the Jodrell Bank 76\,m Mk1 and Mk1a telescopes in the northern hemisphere, and data from the Parkes 64\,m telescope for the southern hemisphere.

The average FWHM of the original map produced by \citet{Haslam1982} is 56\,arcmin. Calibration and zero level offsets were set by an earlier 404\,MHz survey \citep{PaulinyToth1962}. The expected calibration uncertainty is 10\,per\,cent. However, no deconvolution of the beam is possible, therefore we do not know how the calibration changes with angular scale. 

The original 408\,MHz map has in recent years been improved in \citet{Remazeilles2015}, which reduces striping due to instrumental systematics in the original data via Fourier filtering, and also provides an improved subtraction of point sources in the map. For this analysis we use the destriped and desourced map provided by \citet{Remazeilles2015} from the Microwave Background Data Analysis (LAMBDA) website\footnote{\url{http://lambda.gsfc.nasa.gov}}. 

\subsection{Free-Free Template}\label{sec:halphatemplate}

The best currently available method for tracing free-free emission at mid-to-high Galactic latitudes is to use the H$\alpha$ ($\lambda 656.28$\,nm) emission line. This is because for a given {H}{\sc ii} region the brightness at radio frequencies and the intensity of the H$\alpha$ transition are both dependent on just the emission measure and the electron temperature of the plasma \citep{Draine2011}.

H$\alpha$ emission is particularly sensitive to dust extinction, which acts to reduce the H$\alpha$ intensity along any given line of sight. H$\alpha$ maps are also contaminated by stellar continuum emission that must be carefully subtracted. We use the all-sky composite H$\alpha$ maps given by \citet{Dickinson2003} and \citet{Finkbeiner2003} as templates for the free-free emission. Both combine publicly available H$\alpha$ datasets but approach the correction for dust absorption and continuum removal slightly differently. The \citet{Finkbeiner2003} dataset assumes no correction for dust absorption, while for the \citet{Dickinson2003} H$\alpha$ maps we use two dust mixing fractions: $f_\mathrm{d} = 0$ and $f_\mathrm{d} = 0.33$.

\subsection{Dust Templates}\label{sec:dusttemplates}

We use the reprocessed IRAS 100\,$\mu$m data (IRIS), which has a global calibration uncertainty of 13.5\,per\,cent \citep{Miville-Deschenes2005}. The IRAS 100\,$\mu$m ($I_{100\mu\mathrm{m}}$) is primarily a tracer of cold interstellar dust, however it is somewhat sensitive to the local interstellar radiation field (ISRF) since it is near the peak of the thermal dust emission spectrum \citep{Tibbs2013}. 

To trace the cold thermal dust component on the Rayleigh-Jeans tail of the thermal dust spectrum we use the \textit{Planck} 353\,GHz map ($I_{353}$) \citep{planck2016-l01}, as well as  the derived dust optical depth at 353\,GHz ($\tau_{353}$) \citep{planck2014-a12}. These maps are directly proportional to the dust column \citep{planck2013-p06b} but should be insensitive to changes in the ISRF. At frequencies around the peak of the thermal dust spectrum the $\tau_{353}$ data are a better tracer of Galactic dust emission than $I_{353}$ since they do not contain cosmic infrared background anisotropies.

We use the \textit{Planck}-derived dust radiance map \citep{planck2013-p06b}, which is the integrated bolometric intensity of the dust grains. The dust radiance is proportional to the amount of light absorbed by the dust and as such is directly proportional to the ISRF as well as the dust column. 

Finally, to make comparisons with previous works, we also include the 94\,GHz dust brightness map derived from model 8 in \citet{Schlegel1998}, which we will refer to as FDS8. The FDS8 map is an extrapolation of the 100 and 240\,$\mu$m IRAS maps calibrated to the \textit{COBE}-FIRAS spectral data. For many previous studies of dust correlated AME this map has been used, however it has been superseded by direct observations of these frequencies by the {\it Planck} mission.

\subsection{WMAP data}

We use the Wilkinson Microwave Anisotropy Probe (WMAP) 9-year data release \citep{Bennett2013} convolved to a Gaussian $1^{\circ}$ resolution  beam obtained from the LAMBDA website. The WMAP  satellite observations were made using ten differencing assemblies at five frequencies from 23\,GHz to 94\,GHz. The original FWHM resolution of the instrument was $0\pdeg93$ to $0\pdeg23$. For the calibration uncertainty of the WMAP maps we assign a 3\,per\,cent uncertainty to account for colour corrections and residual beam asymmetries. We convert all the maps to brightness temperature units from thermodynamic units relative to the CMB.

\subsection{\textit{Planck} data}

We use the 2018 release of the \textit{Planck} all-sky maps \citep{planck2016-l01}. The \textit{Planck}  low-frequency instrument (LFI)  was a pseudo-correlation radiometer that operated between 30 and 70\,GHz, with FWHM resolutions of 33 to 13\,arcmin; and the high-frequency instrument (HFI) used bolometers that spanned a frequency range of 100 to 857\,GHz, with resolutions of 7.2 to 4.5\,arcmin FWHM. We adopt calibration uncertainties for the LFI of 3\,per\,cent \citep{planck2013-XV} and HFI of 5\,per\,cent  \citep{planck2016-l01}; these uncertainties account for colour corrections, residual beam asymmetries, and other systematics. Units were converted from thermodynamic units relative to the CMB to brightness temperature units.

\vspace{-0.1cm}
\section{Template Fitting}\label{sec:templatefitting}

\subsection{Method}\label{sec:method}

To estimate the contribution of a given emission component to each frequency map we use a template fitting method (also referred to as a correlation analysis); a technique that has been well established for studying Galactic foregrounds~\citep{Kogut1996b,deOliveira-Costa1997,Davies2006,planck2011-7.3,Peel2012,Ghosh2012,planck2014-XXII}. The method assumes that the sky signal can be decomposed into a linear combination of $N_Z$ template maps, where each template is chosen to trace a single emission component. Template fitting is performed by solving for the template coefficients $\mathbf{a}$ in
\begin{equation}\label{eqn:template1}
    \mathbf{d} = \mathbf{Z}\mathbf{a} + n,
\end{equation}
where each column of $\mathbf{Z}$ ($N_\mathrm{pix} \times N_Z$) contains the $N_\mathrm{pix}$ pixel intensities of the templates used to decompose the sky vector $\mathbf{d}$ ($N_\mathrm{pix}$). We want to solve for the template coefficients vector $\mathbf{a}$, where each coefficient describes the radio brightness of a given emission component at a given frequency per unit template. Least-squares solution gives
\begin{equation}\label{eqn:template2}
    \hat{\mathbf{a}} = \left(\mathbf{Z}^T\mathbf{N}^{-1}\mathbf{Z}\right)^{-1} \mathbf{Z}^{T}\mathbf{N}^{-1}\mathbf{d},
\end{equation}
where $\mathbf{N}^{-1}$ is the pixel noise covariance matrix. The details of the noise covariance matrix are described in \autoref{sec:noise}. 

We fit for the synchrotron, dust, and free-free emission components, as well as an arbitrary offset. Before we perform the template fit we first subtract the mean offset of each template. Subtracting an offset from each template does not impact the fitted coefficients but does improve convergence and removes arbitrary correlations between templates and the data.

Coefficient uncertainties can be calculated using the covariance of the coefficients as
\begin{equation}\label{eqn:boot1}
    C_a = \mathbf{Z}^T\mathbf{N}^{-1}\mathbf{Z}~,
\end{equation}
where each parameter is as defined in \autoref{eqn:template2}. However, estimating the coefficient uncertainties in this manner requires both the templates to be perfect representations of the underlying emission, and for the noise to be Gaussian distributed in each dataset. Therefore we instead estimate uncertainties and the correlation between coefficients using the bootstrapping method \citep{Efron1979,Efron1986}. The bootstrapping method gives unbiased estimates of the uncertainties in the coefficients by randomly resampling with replacement the pixels in each region. Resampling is done 1000\,times for each region resulting in uncertainties in the bootstrapped coefficient uncertainties of $\approx 3$\,per\,cent. We estimate the coefficient covariance matrix by  averaging over the outer-product of all the estimates of $\mathbf{a}$,
\begin{equation}\label{eqn:boot2}
    C_a = \left<\hat{\mathbf{a}}\hat{\mathbf{a}}^T\right>.
\end{equation}
We find that for regions far from bright sources of Galactic emission the differences between the coefficient uncertainties estimated using \autoref{eqn:boot1} and \autoref{eqn:boot2} were small---the bootstrapped uncertainties were $10$--$20$\,per\,cent larger for data between $5$ and $60$\,GHz. However, for regions nearer to the Galactic plane, or coincident with bright features (i.e., Eridanus/Orion---regions 69, 82, 83, 97 in \autoref{fig:regions1}) the bootstrapped uncertainties can be as much as an order-of-magnitude larger. For the higher frequencies ($\nu > 143$\,GHz), the bootstrapped uncertainties were systematically larger in all regions by as much as an order-of-magnitude, which is likely due to not including any additional sources of noise at high frequency other than the instrument noise. Therefore it is clear that the bootstrapped uncertainties are more reliable and representative of the data.

\subsection{Free-Free Emission Removal}\label{sec:ffremove}

The \mbox{C-BASS} map contains contributions from both synchrotron and free-free emission, the latter from the warm ionized medium (WIM). When using the 4.76\,GHz \mbox{C-BASS} map as a synchrotron template we first subtract a global estimate of the free-free emission using the H$\alpha$ data. We use the best-fitting H$\alpha$ coefficient at 4.76\,GHz  of $T^\mathrm{ff}_{b}/I_{\mathrm{H}\alpha} = (195 \pm 5)$\,$\mu$K/R (see \autoref{sec:halpha} for derivation of this value) to scale the H$\alpha$ data, and then subtract this from the \mbox{C-BASS} map.

It is possible that subtracting the free-free emission in the \mbox{C-BASS} map may result in systematic biases in later results (since we are using the same H$\alpha$ map to trace free-free emission at higher frequencies). Therefore we subtracted free-free templates from the \mbox{C-BASS} data using fixed electron temperature values of 5000, 6000, 7000, and 8000\,K using \autoref{eqn:freefree1}. Changing the amplitude of the subtracted free-free component resulted in no significant change in the fitted coefficients discussed in \autoref{sec:results}, which is not unexpected since free-free emission contributes less than 20\,per\,cent of the total emission at 4.76\,GHz at high Galactic latitudes.

\subsection{Mask}\label{sec:mask}

\begin{figure}
  \centering
\includegraphics[width=0.45\textwidth]{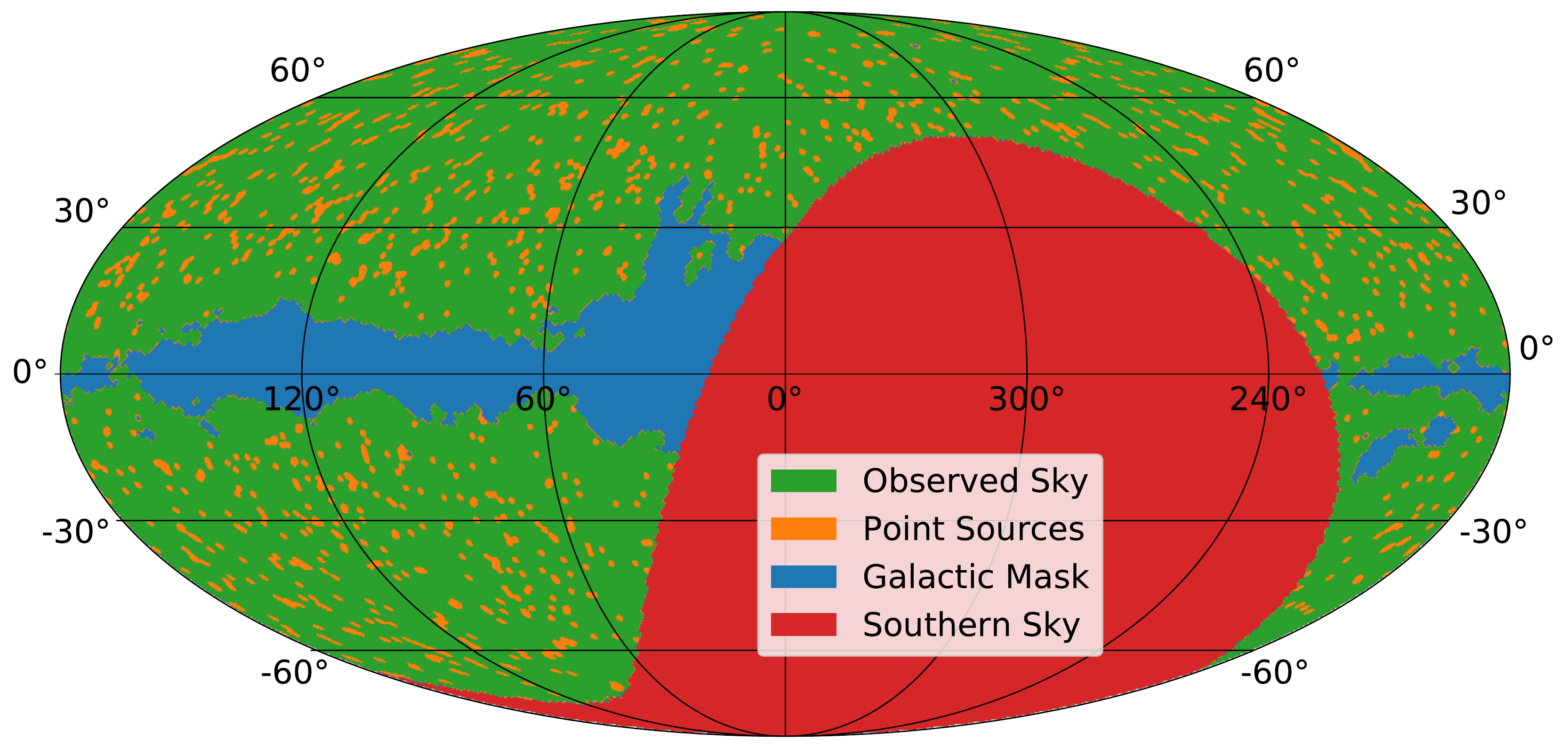}
\caption{Sky mask used for template fitting plotted in a Galactic Mollweide projection. Only the \textit{green} pixels are used in the template fitting analysis.} 
\label{fig:masks}
\end{figure}

Template fitting is only as effective as the templates that trace the underlying emission components at a given frequency. In order to optimize the template fitting procedure we mask regions of the sky that lie along the plane of the Galaxy where components are highly correlated at $1^\circ$ scales or larger and the H$\alpha$ template suffers from significant extinction, making the separation of components unreliable. The Galactic plane is masked by first median filtering the C-BASS map on $5^{\circ}$ scales, and then masking the brightest 10\,per\,cent of pixels in the filtered map. The Galactic plane mask is shown as the \textit{blue} region in \autoref{fig:masks}.

Point sources can also be problematic for template fitting as they can dominate over the diffuse background emission, resulting in a highly biased estimate of the amplitudes describing the diffuse components. At lower frequencies we use the source-subtracted 408\,MHz and 4.76\,GHz \mbox{C-BASS} maps, thus we only mask the very brightest sources or sources that are found to have large residuals after subtraction. \autoref{fig:masks} shows bright sources ($>10$\,Jy at 4.76\,GHz; \citealt{Grumitt2020}) masked by a $2^{\circ}$ diameter aperture. There are also some fainter point sources ($S_\nu < 10$\,Jy at 4.76\,GHz) that leave residuals in the \mbox{C-BASS} map after subtraction; we mask these using a smaller $1^{\circ}$ diameter aperture.

At WMAP and \textit{Planck} frequencies the maps are not confusion-limited since the majority of background radio sources have  steep spectra and are generally too faint to detect in the WMAP/\textit{Planck} data. However, there are still a large number of flat-spectrum sources present at high frequencies, and we mask these using  the 30\,GHz \textit{Planck} catalogue of compact sources (PCCS) \citep{planck2014-a35}.  We mask 893 PCCS sources in the range $50 < S < 1000$\,mJy with an aperture of diameter $1\pdeg5$, and 628 sources with flux densities $S > 1$\,Jy with a 3$^{\circ}$ wide aperture. Both the PCCS sources and the \mbox{C-BASS} sources described earlier are shown in \autoref{fig:masks} as the \textit{orange} regions. 

\subsection{Regions}\label{sec:regions}

\begin{figure}
    \centering
    \includegraphics[width=0.45\textwidth]{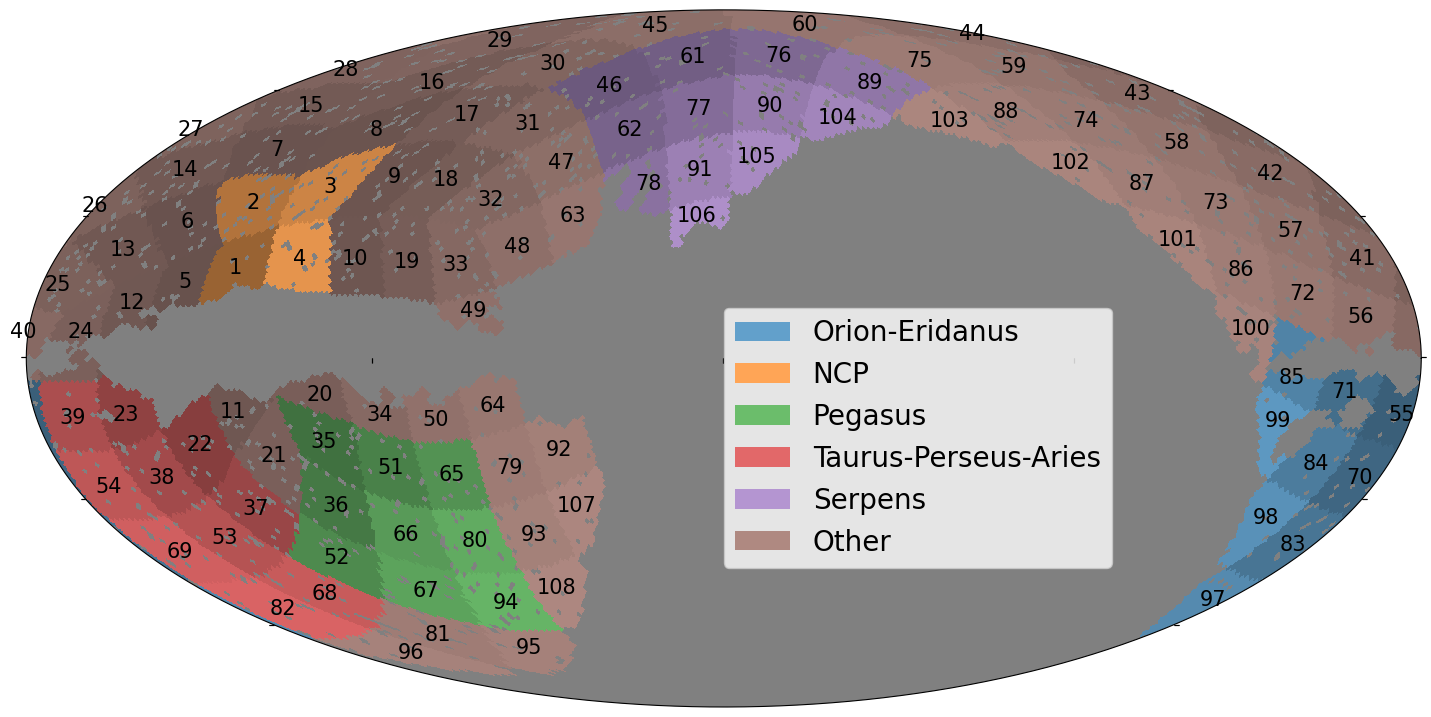}
    \includegraphics[width=0.45\textwidth]{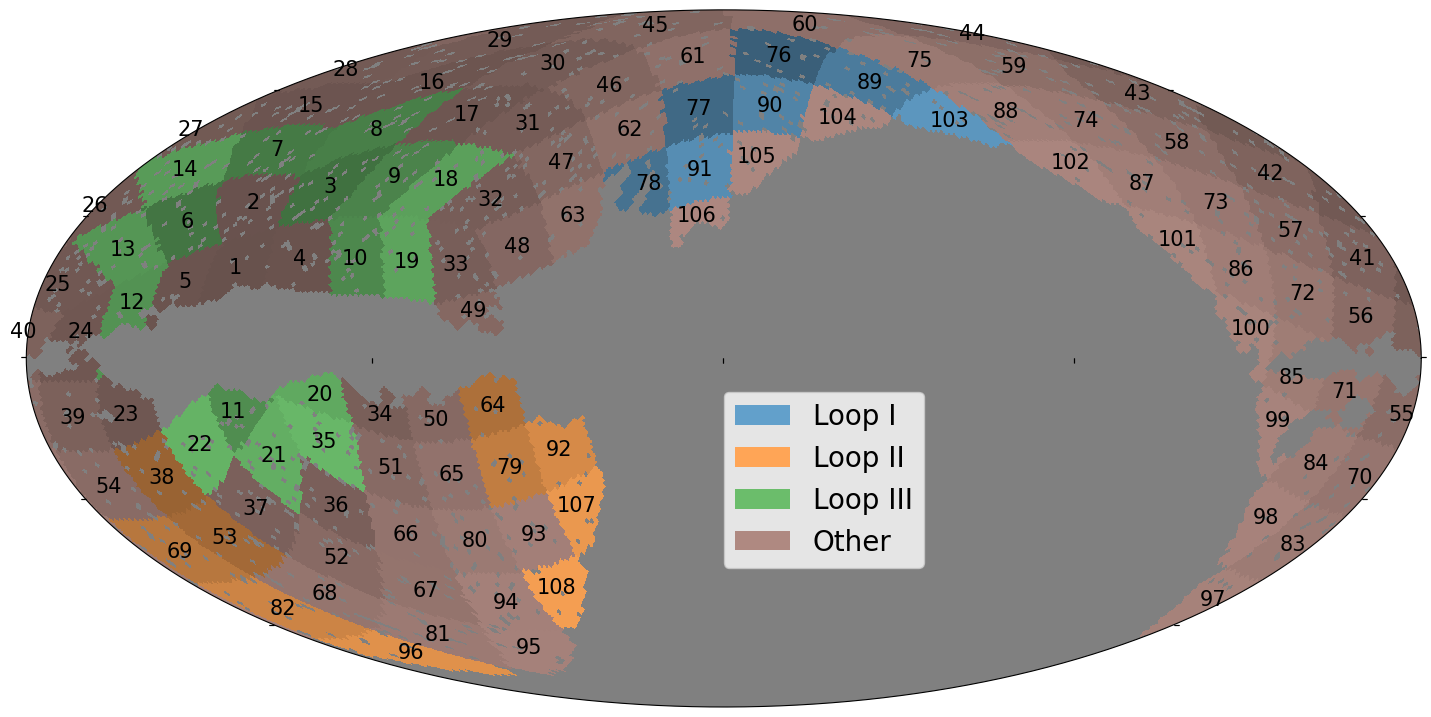}
    \caption{Mollweide projection in Galactic coordinates of the 108 $N_{\rm side}=4$ regions used to sub-divide the sky and their numerical identifiers. \textit{Top}: Region group designations discussed in \autoref{sec:dust}. \textit{Bottom}: Region groups of synchrotron loops discussed in \autoref{sec:sync}.}
    \label{fig:regions1}
\end{figure}

For this analysis we divided the sky into equal sized regions using the \texttt{Healpix} grid \citep{gorski2005} on the celestial sphere. We chose to use a region size of $N_\mathrm{side}=4$, which equates to regions with areas of approximately 200\,deg$^2$. There are a total of 108 regions, which are shown in \autoref{fig:regions1}. The maximum number of $N_\mathrm{side}=64$ pixels within a region is 256, but after masking the number of pixels within each region ranges between 104 and 252 (regions with less than 100 pixels are excluded). The choice of region size is important because regions that are too small have larger spatial correlations between emission types as the template fitting method relies on there being spatial differences between components to work. Whereas larger sized regions become more susceptible to systematics since large-scales are generally harder to constrain. Further, larger regions result in less information on the spatial variations across the sky. We found that a region size of $N_\mathrm{side}=4$ was a good balance between these considerations given the resolution of \mbox{C-BASS} data.

\subsection{Noise Covariances}\label{sec:noise}

\begin{figure}
  \centering
\includegraphics[width=0.4\textwidth]{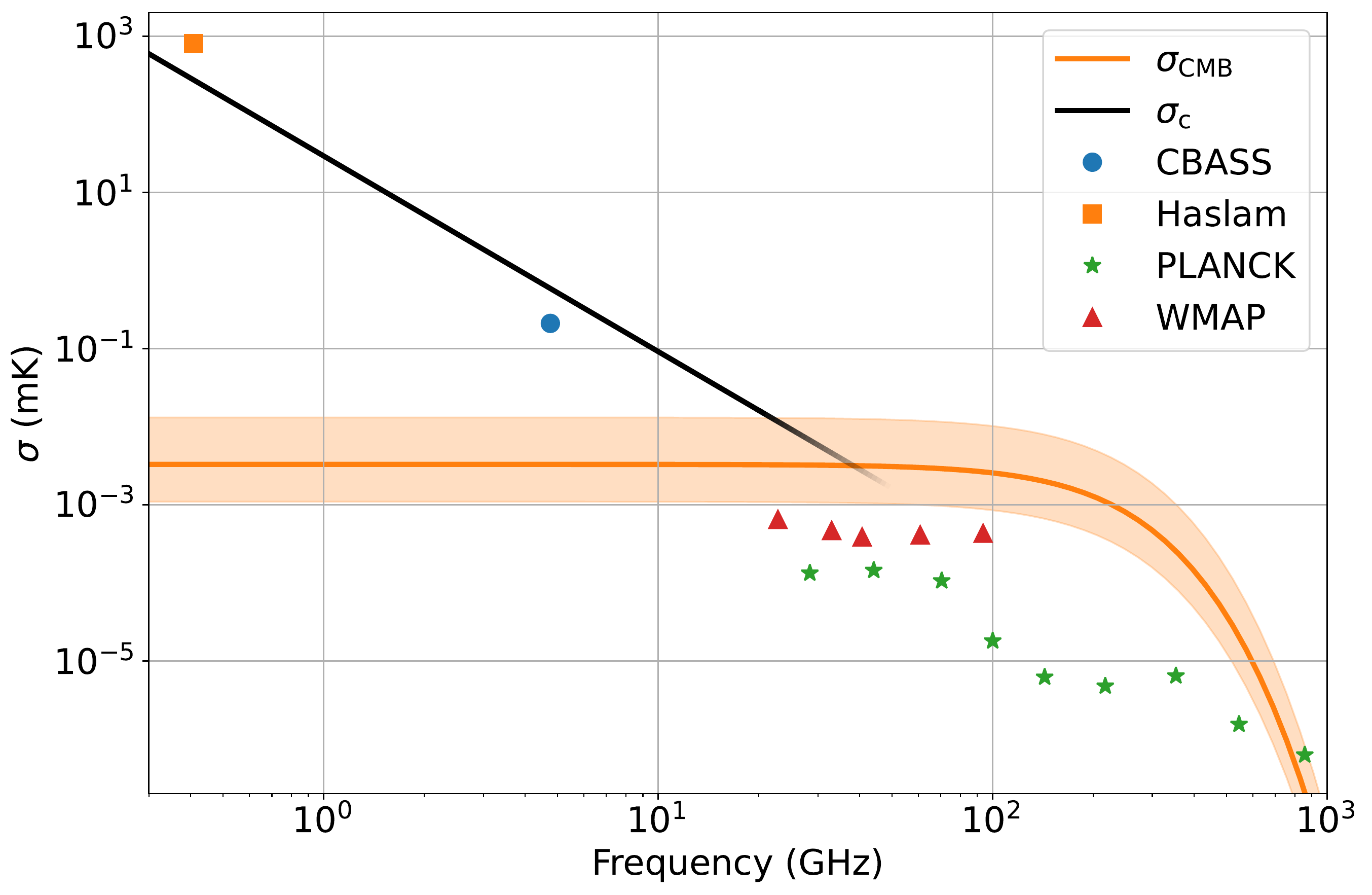} 
\caption{Summary of the mean instrumental pixel noise RMS for each dataset (\textit{points}). The \textit{black} line is the confusion noise model given by \autoref{eqn:conf_model}, we fade the line at high frequencies as at these frequencies the steep-spectrum background sources are less relevant. The \textit{orange} bounded region shows the range of the residual CMB fluctuations for pixels of approximately 1\,deg$^2$ ($N_\mathrm{side}=64$).}
\label{fig:bkgd_noise}
\end{figure}

As described in \autoref{sec:method} we estimate the uncertainties in the fitted coefficients using bootstrapping, and not the intrinsic noise covariances of each map. However, we still need to calculate the noise covariance to correctly weight each pixel. \autoref{fig:bkgd_noise} provides a summary of all the different contributions to the total noise budget at each frequency. We can see from the figure that the instrumental noise is never the dominant source of noise in the maps---except for the 408\,MHz data. At higher frequencies we have neglected contributions to the noise due to confused infrared galaxies. We give a more detailed description of each contribution below.

For this analysis we assume the covariance matrix $\mathbf{N}$ is diagonal and is the sum of contributions from instrumental noise, confused background radio sources, and residual CMB fluctuations. As mentioned in \autoref{sec:data}, the SMICA estimate of the CMB \citep{planck2013-p06} was subtracted from all the maps, therefore we only consider \textit{uncorrelated} CMB residual fluctuations. The noise covariance matrix is therefore constructed as
\begin{equation}
    \mathbf{N} = \mathbf{N}_\mathrm{noise} + \mathbf{N}_{\delta\mathrm{CMB}} + \mathbf{N}_\mathrm{PS},
\end{equation}
where $\mathbf{N}_\mathrm{noise}$ describes noise contributions from the instruments, $\mathbf{N}_{\delta\mathrm{CMB}}$ due to uncorrelated errors in the CMB subtraction, and $\mathbf{N}_\mathrm{PS}$ due to confusion from background radio sources.

Instrument noise covariances are calculated for WMAP using the maps of integration time per pixel (hit maps) and the receiver sensitivities provided in \citet{Bennett2013}. For \textit{Planck} and \mbox{C-BASS}, variance maps are calculated during the map-making stage for each pixel and can be used directly. The noise variances for all three surveys are corrected for smoothing of the noise by rescaling the noise by the integral of the Gaussian beam transfer function. For the  408\,MHz map there is no information on the per pixel noise in the map, however \citet{Remazeilles2015}  estimated the noise at $N_\mathrm{side} = 512$ (i.e., $\approx 7$\,arcmin size pixels) to be $\approx 800$\,mK, which is equivalent to 0.1\,mK\,deg$^{-1}$.

To estimate the noise due to the randomly (Poisson) distributed background radio sources we used the compilation of differential source counts for frequencies of 0.1--1000\,GHz provided in \citet{Zotti2010}. We calculated the power spectrum  of the sources by integrating differential source counts
\begin{equation}\label{eqn:conf_power}
   C_\ell = \left(\frac{2 k \nu^2}{c^2}\right)^{-2} \int_{S_\mathrm{min}}^{S_\mathrm{max}} S^2 \left(\frac{\mathrm{d}N}{\mathrm{d}S}\right) \mathrm{d}S ,
\end{equation}
where $\frac{\mathrm{d}N}{\mathrm{d}S}$ is the model of the differential source counts at a given frequency, $k$ is the Boltzmann constant, $c$ is the speed of light, and $\nu$ is the observing frequency. We use an upper limit of $S_\mathrm{max} = 1$\,Jy to match the flux density at which we mask sources, and a lower limit ($S_\mathrm{min}$) that matched each source catalogue's minimum flux density. To estimate the confusion noise in the map we then multiply the background source power spectrum from \autoref{eqn:conf_power} by a Gaussian beam transfer function ($\mathrm{FWHM}=1^\circ$) and integrate over all $\ell$ as
\begin{equation}
  \sigma_c^2 = \frac{\sum_\ell\left(2\ell + 1\right) C_\ell B_\ell}{4 \pi},
\end{equation}
where $B_\ell$ is the power of the Gaussian beam transfer function. We then fitted a power-law to $\sigma_c^2$ measured at each frequency between $1$ and $20$\,GHz to get a model of the confusion noise due to steep-spectrum radio sources
\begin{equation}\label{eqn:conf_model}
  \log_{10}\left(\frac{\sigma_c^2}{\mathrm{K}^2}\right) = -5.0 \log_{10}\left(\frac{\nu}{\mathrm{GHz}}\right) - 3.1 + \log_{10}\left(\frac{\theta^2}{\mathrm{deg}^2}\right),
\end{equation}
where $\theta$ is the beam FWHM, and $\nu$ is the frequency in GHz. \autoref{eqn:conf_model} is only an applicable model of confusion up to a few tens of gigahertz, after which the background source population is dominated by flatter-spectrum radio sources \citep[e.g.,][]{Healey2007}, and at infrared frequencies the cosmic infrared background becomes the dominant source of confusion \citep[e.g.,][]{Guiderdoni1997}. These additional sources of uncertainty are accounted for by the bootstrapping method discussed in \autoref{sec:templatefitting}.

In \autoref{fig:bkgd_noise} the contribution of the residual CMB noise exceeds the noise in the data for most of the WMAP and \textit{Planck} frequencies. To calculate the contribution of the residual CMB noise ($\mathbf{N}_{\delta\mathrm{CMB}}$) we use the five CMB maps published in \citet{planck2013-p06}. We calculated each unique difference pair between these maps resulting in a cube of ten CMB difference maps. We downsampled each difference map from its original resolution  ($N_\mathrm{side}=2048$) to the same pixel size as the rest of the data ($N_\mathrm{side}=64$), and stored the variance within each downsampled pixel. We then took the CMB noise contribution to be the mean variance across all of the 108 regions. 

\vspace{-0.1cm}
\section{Template Fitting Results}\label{sec:results}

\subsection{Dust}\label{sec:dust}

Here we present the template fitting coefficients derived using the dust templates. There are a number of different dust tracers that are available. In this section we will be comparing five dust tracers (\autoref{tab:summary}): The \citet{Finkbeiner1999} model-8 94\,GHz predicted dust brightness map (hereafter FDS8); the IRAS 100\,$\mu$m map \citep[$I_{100\mu\mathrm{m}}$;][]{Miville-Deschenes2005}; the \textit{Planck} integrated dust intensity or dust radiance map \citep[$\mathcal{R}$;][]{planck2013-p06b}; the \textit{Planck} derived thermal dust optical depth at 353\,GHz \citep[$\tau_{353}$;][]{planck2013-p06b}; and the \textit{Planck} 353\,GHz intensity data \citep[$I_{353}$;][]{planck2016-l01}. We use these five dust tracers in this section to see which of them best traces AME, to determine whether AME can be explained by a hard synchrotron component, and to study the distribution of AME emissivity across the sky. Later, in \autoref{sec:fitting}, we fit spectral models to the dust coefficients presented in this section.

\begin{figure}
    \centering
    \includegraphics[width=0.4\textwidth]{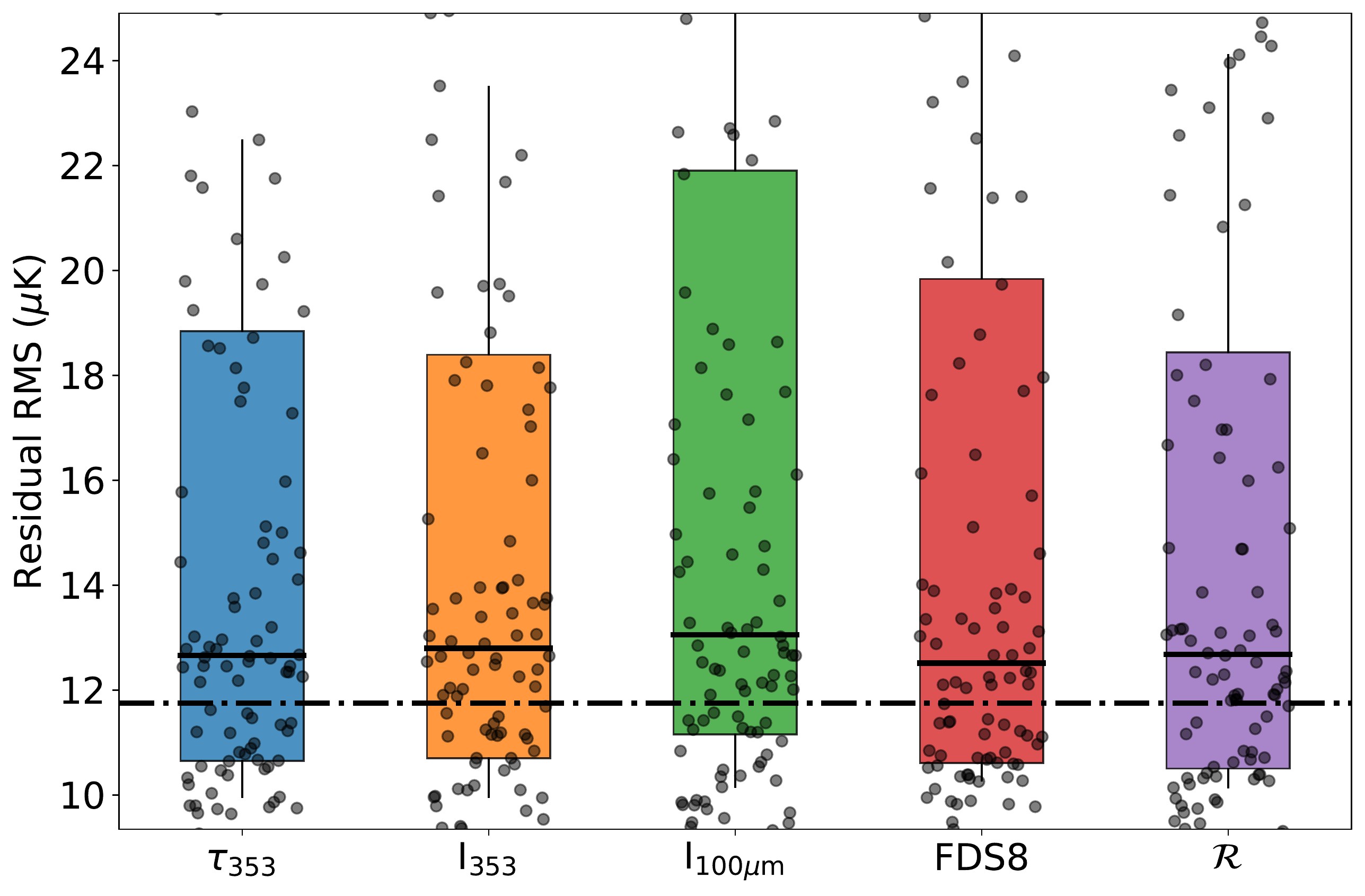}
    \caption{Box-plot of the residual RMS at 22.8\,GHz for each region and each dust template. The coloured boxes represent the first and third quartiles of each distribution, and the \textit{bold-black} line is the median of each distribution. The dot-dash black line indicates the median expected RMS at 22.8\,GHz.}
    \label{fig:dust1_k}
\end{figure}

One of the biggest challenges for AME studies has been determining which dust map best traces AME. In order to compare the five different dust tracers we look at the residual RMS at 22.8\,GHz in each region after subtracting our best model for the emission. In the case that the emission is perfectly modelled we expect that the noise in the map will be equivalent to a combination of instrumental, confusion, and residual CMB noise. In \autoref{fig:dust1_k} we show the RMS, at 22.8\,GHz, of each region after subtracting the best-fitting dust, synchrotron, and free-free templates. We did this for each dust template and find that there is no global preference for any one dust tracer over another on the scales of the region sizes ($\approx 10^\circ$). It is not surprising that we find little difference between the dust tracers since they are all tracing features at high latitudes that are spatially nearby \citep{Green2019} and tend to share similar environmental conditions \citep{planck2013-p06b}.

Another long-standing challenge with studies of AME on large scales has been separating it from synchrotron emission \citep[e.g.,][]{Gold2011,planck2014-a31}. AME has a peaked spectrum that can be broadly described by a log-normal distribution centred on 20--30\,GHz \citep[e.g.,][]{Bonaldi2011,planck2013-XII,Stevenson2014,Cepeda-Arroita2021}. However, at frequencies above the spectral peak, AME can be described reasonably well by just a simple steep-spectrum power-law, which can be challenging to separate from the synchrotron spectrum. This has led to the suggestion that Galactic synchrotron emission could be decomposed into two broad synchrotron components. The first being a soft/steep spectrum synchrotron component which is what is observed in the 408\,MHz all-sky data, and the second, describing AME,  is a dust-correlated hard/flat synchrotron component associated with star formation regions \citep{Bennett2003,Davies2006}.

\begin{figure}
    \centering
    \includegraphics[width=0.25\textwidth]{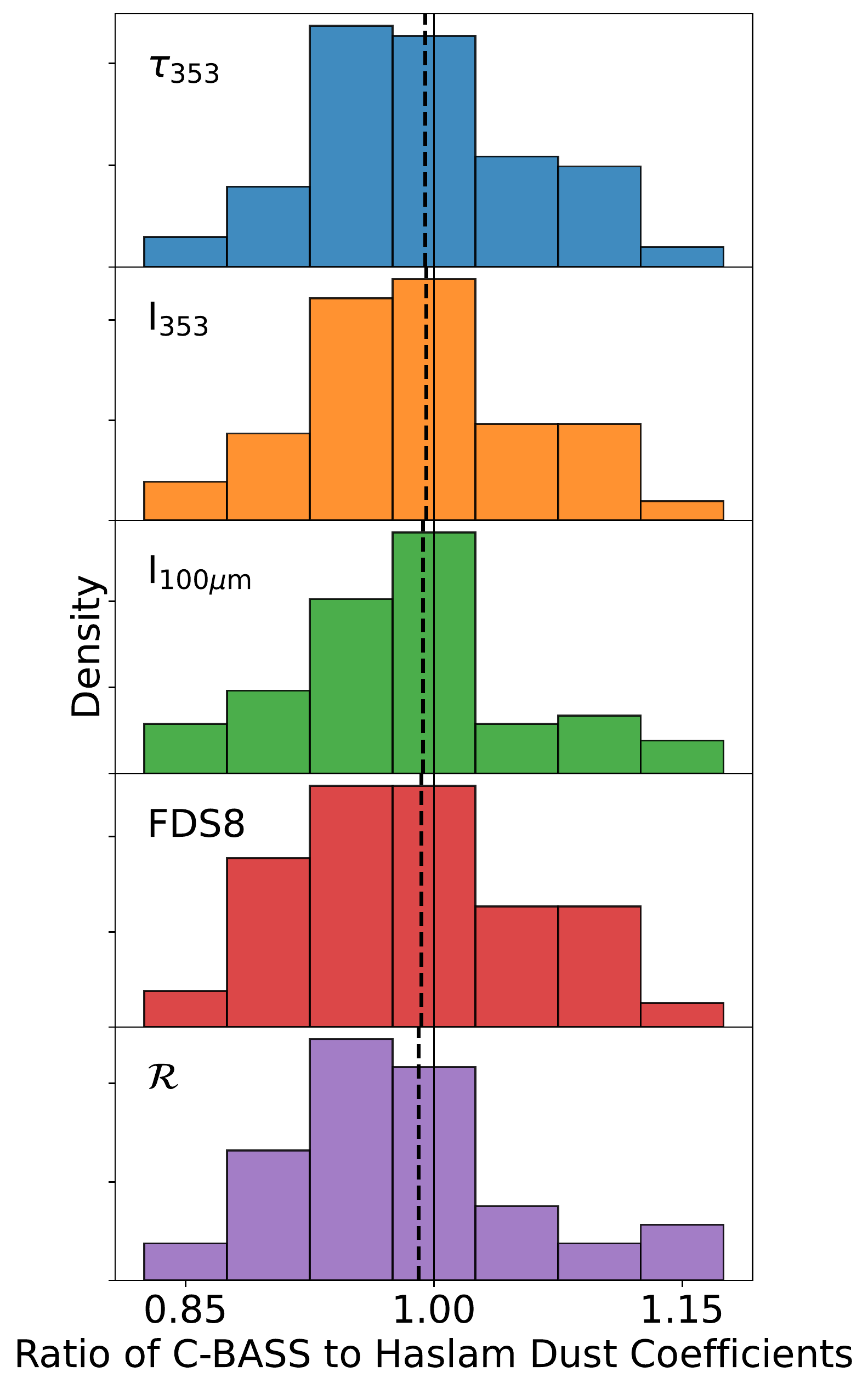}
    \caption{Weighted histograms of the fitted dust coefficients when using the \mbox{C-BASS} 4.76\,GHz template over those fitted when using 408\,MHz Haslam synchrotron template for each dust tracer at 22.8\,GHz. The \textit{dashed} lines indicate the weighted mean for each dust template. The weighted average ratio is consistent with unity for each dust template.}
    \label{fig:dust1_j}
\end{figure}

If AME is due to a hard/flat spectrum synchrotron component we would expect to find AME in the 4.76\,GHz \mbox{C-BASS} data. If this were the case then previous estimates of the dust coefficients that used the 408\,MHz synchrotron template to trace synchrotron emission  around 20--60\,GHz would be significantly different to those reported here due to correlations between the dust and 4.76\,GHz synchrotron templates. These differences can be either  positive or negative depending on the local relative brightnesses of the synchrotron emission and potential AME at 4.76\,GHz. In \autoref{fig:dust1_j} we show a weighted histogram of the ratios of the dust coefficients measured using the 4.76\,GHz synchrotron template over the 408\,MHz template for each dust tracer (excluding a single outlier region---discussed below). We find that the weighted mean ratio is consistent with unity for each template. We find that most individual regions are also consistent, within uncertainties, with no change in the emissivity of AME when changing the synchrotron tracer. We therefore find no evidence that AME is a flat-spectrum synchrotron component.

We find that region 46 (Serpens group---\autoref{fig:regions1}) is the only region to have a significant ($>5\sigma$) change in the ratio of the dust coefficients. Region 46 lies along the northern edge of the North Polar Spur (NPS) and contains both bright diffuse synchrotron and diffuse dust emission. When the synchrotron component is fitted using the 4.76\,GHz template the dust coefficient is \mbox{$(19.5\pm1.4)$\,$\tau_{353}$/K}, which is more than three times greater than the equivalent when using the 408\,MHz template of \mbox{$(5.9\pm1.5)$\,$\tau_{353}$/K}. We find that there is a negative correlation between the dust and synchrotron coefficients in this region when using either the 4.76\,GHz template ($\rho_{4.76} = -0.58$) or the 408\,MHz template ($\rho_{408} = -0.48$), indicating that the pixel brightness distributions in the 408\,MHz, 4.76\,GHz, and $\tau_{353}$ templates are equally correlated. The measured difference in dust emissivity could be due to region selection effects. To test this we looked at the region again with a larger grid size ($N_\mathrm{side}=2$, or a region of approximately 800\,deg$^2$) and found that the dust emissivity measured using the 408\,MHz template of $(17.6\pm1.2)$\,$\tau_{353}$/K is more consistent with the dust emissivity using the 4.76\,GHz template of $(19.6\pm1.0)$\,$\tau_{353}$/K. The reason for the difference between $N_\mathrm{side}=2$ and $N_\mathrm{side}=4$ when using the 408\,MHz template is possibly due to the data processing of the 408\,MHz map \citep{Remazeilles2015} that averages out over a larger region.

\begin{table*}
\caption{Median dust coefficients for the \mbox{C-BASS} 4.76\,GHz, WMAP 22.8\,GHz and {\it Planck} 28.4\,GHz data for each dust template. The synchrotron component is traced using \mbox{C-BASS} for the WMAP and \textit{Planck} data, and is traced using the 408\,MHz map in the \mbox{C-BASS} data. The 25$^{th}$ and 75$^{th}$ percentiles are given as subscripts and superscripts, respectively. The uncertainty given includes both the statistical and calibration uncertainties. For comparison, we include both the weighted average of the regions and the best-fit to all regions simultaneously.}
\label{tab:dust1_a}%
\renewcommand{\arraystretch}{1.4}
\centering%
\begin{tabular}{l|c|c|c|c|c|c|c}%
\hline%
\hline%
&\multicolumn{2}{c|}{C-BASS 4.76\,GHz}&\multicolumn{2}{c|}{WMAP 22.8\,GHz}&\multicolumn{2}{c|}{Planck 28.4\,GHz}&\\%
\hline%
&$\left< a \right>_\mathrm{regions}$&$a_\mathrm{all}$&$\left< a \right>_\mathrm{regions}$&$a_\mathrm{all}$&$\left< a \right>_\mathrm{regions}$&$a_\mathrm{all}$&Unit\\%
\hline%
$\tau_{353}$&$3^{~124}_{-86}\pm 2$ & $61\pm3$&11.5$^{15.7}_{10.0}$$\pm$0.4&10.0$\pm$0.3&5.8$^{8.7}_{5.3}$$\pm$0.2&5.3$\pm$0.2&K/$\tau_{353}$\\%
$\mathrm{I}_{353}$&0.2$^{~10.1}_{-5.0}$$\pm$0.2&5.7$\pm$0.2&1.01$^{1.29}_{0.87}$$\pm$0.03&0.94$\pm$0.03&0.52$^{0.69}_{0.43}$$\pm$0.02&0.5$\pm$0.02&K/K\\%
$\mathrm{I}_{100\mu \mathrm{m}}$&43$^{~320}_{-90}$$\pm$7&178$\pm$9&26.3$^{32.7}_{20.6}$$\pm$0.8&35.9$\pm$1.1&14.0$^{17.7}_{10.7}$$\pm$0.4&19.3$\pm$0.6&$\mu$K/$\mathrm{MJy} \mathrm{~sr}^{-1}$\\%
FDS8&1.6$^{~80.3}_{-41.4}$$\pm$1.2&36.5$\pm$1.6&6.9$^{8.6}_{5.9}$$\pm$0.2&6.4$\pm$0.2&3.5$^{4.6}_{2.9}$$\pm$0.1&3.4$\pm$0.1&K/K\\%
$\mathcal{R}$&600$^{~7000}_{-2700}$$\pm$130&3680$\pm$160&600$^{710}_{480}$$\pm$20&710$\pm$20&310$^{380}_{260}$$\pm$10&380$\pm$12&K/($\mathrm{W} \mathrm{m}^{-2} \mathrm{ ~sr}^{-1}$)\\%
\hline%
\end{tabular}%
\end{table*}

In \autoref{tab:dust1_a} we give the measured coefficients for a joint fit to all regions simultaneously ($a_\mathrm{all}$---hereafter \textit{all-region fit}). We note that for the $\mathrm{I}_{353}$ and FDS8 templates there is no significant difference between the all-region and region average coefficients. However, the all-region $\tau_{353}$ estimate is lower by approximately 10\,per\,cent, and the all-sky $\mathrm{I}_{100\mu\mathrm{m}}$ and $\mathcal{R}$ fit are systematically higher by almost 50\,per\,cent. The fact that we do not find all dust tracers to be systematically higher or lower than the region averages suggests that this is a real difference between the five dust tracers. This could occur because the all-region fit is sensitive to larger angular scale emission that is effectively filtered out when fitting each region individually. 

Comparing the coefficients presented in \autoref{tab:dust1_a} to similar analyses in the literature that used the 408\,MHz map to trace Galactic synchrotron emission, we find our results are consistent, again suggesting flat-spectrum synchrotron is not a significant component at frequencies of 5\,GHz and higher. For the FDS8 template, previous measurements found the average coefficient at high latitudes to be $(6.6\pm0.3)$\,$\mu$K/mK at 22.8\,GHz \citep[Kp2 mask;][]{Davies2006}. There are also several papers that look at $\tau_{353}$, for which, at high latitudes, \citet{Hensley2016} found the coefficient to be $(7.9\pm2.6)$\,K/$\tau_{353}$ at 30\,GHz, slightly larger than the average we find here but within the interquartile range of the values. At 22.8\,GHz \citet{planck2014-a31} finds $\tau_{353}$ to be slightly lower than we find it; $(9.7\pm1.0)$\,K/$\tau_{353}$ at high latitudes. Although our results are consistent with most previous measurements we do find that the dust coefficient we measure using $\mathcal{R}$ is 30--40\,per\,cent lower than that in \citet{Hensley2016}. We believe this discrepancy is due to the \commander-derived AME map, which is known to be discrepant with other AME measurements at the 30--50\,per\,cent level \citep{planck2016-l04,planck2014-a31,Cepeda-Arroita2021}. \autoref{tab:all_coeff1} and \autoref{tab:all_coeff2} provide the synchrotron, dust, and free-free coefficients at 22.8\,GHz for all regions using the 4.76\,GHz, $\tau_{353}$, and H$\alpha$ ($f_D=0.33$) templates.

\begin{figure*}
    \centering
    \includegraphics[width=0.6\textwidth]{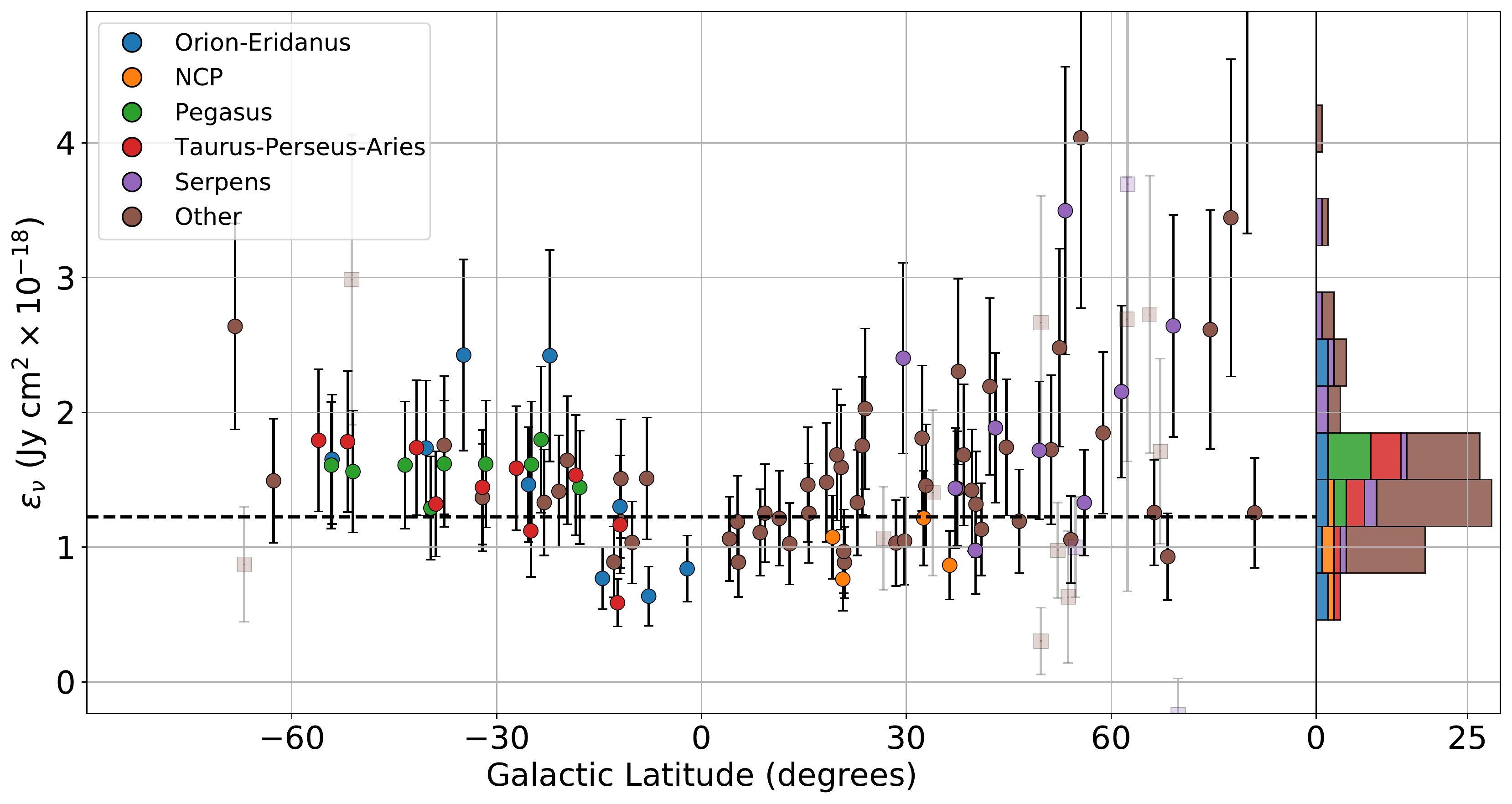}
    \caption{Region-to-region AME emissivity at 22.8\,GHz derived using \autoref{eqn:dust_emiss} and the $\tau_{353}$ dust coefficients. Marker colours indicate the regions associated region group as defined in \autoref{fig:regions1}. The weighted mean AME emissivity at 22.8\,GHz over all regions is shown with the \textit{dashed} black line. Semi-transparent symbols are for regions that have coefficient $\mathrm{SNR}<5$. The \textit{right} panel shows the histogram of the coefficients colour-coded by region group.}
    \label{fig:dust1_i}
\end{figure*}

In \autoref{fig:dust1_i} we show the AME emissivity defined as the intensity per total hydrogen column density for each region. We can derive the emissivity from the fitted $\tau_{353}$ dust parameters by the relation 
\begin{equation}\label{eqn:dust_emiss}
  \epsilon_\nu = 3.1 \left(\frac{a_T}{{\rm K}/\tau_{353}}\right)   \left(\frac{\sigma_{e,353}}{\tau_{353}/\mathrm{H}~\mathrm{cm}^{-2} }\right) \left(\frac{\nu}{\mathrm{GHz}}\right)^2 \times 10^4 \mathrm{\,Jy\,sr}^{-1}\mathrm{\,cm}^{2}\mathrm{H}^{-1},
\end{equation}
where $a_\mathrm{T}$ is the fitted dust coefficient, and the average dust opacity at 353\,GHz is \mbox{$\sigma_{e,353} = (7.0 \pm 2.0) \times 10^{-27}$\,cm$^2$\,H$^{-1}$} for $|b|>15^\circ$ \citep{planck2013-p06b}. The frequency dependence in \autoref{eqn:dust_emiss} is due to the conversion from brightness temperature to flux density per steradian. The colours of the markers in \autoref{fig:dust1_i} indicate which region group each region belongs to, we chose region groups based on known dust features (region group locations are shown in \autoref{fig:regions1}). We found that there is no significant difference in the AME emissivity between these region groups implying that the spinning dust emission origin between region groups is similar.

The mean fitted $\tau_{353}$ dust coefficient is $(11.5\pm0.4)$\,K/$\tau_{353}$ at 22.8\,GHz (\autoref{tab:dust1_a}) which gives a mean AME emissivity at high Galactic latitudes of $(1.3 \pm 0.4) \times 10^{-18}$\,Jy\,sr$^{-1}$\,cm$^{2}$\,H$^{-1}$. The weighted mean emissivity across all regions in \autoref{fig:dust1_i} at 22.8\,GHz is $(1.2 \pm 0.4) \times 10^{-18}$\,Jy\,sr$^{-1}$\,cm$^{2}$. We find the emissivity we measure using \autoref{eqn:dust_emiss} is systematically lower than the theoretical estimates, which at 22.8\,GHz can vary between $(3-12) \times 10^{-18}$\,Jy\,sr$^{-1}$\,cm$^{2}$\,H$^{-1}$ depending on the environment \citep[e.g.,][]{Draine1998a,AliHaimoud2009}. Although these emissivities are lower than initially expected, they are not beyond the possibility of the spinning dust model given the large number of parameters that can be tuned. Alternatively, the low emissivities could indicate a bias between the AME emissivity and the dust column traced by the $\tau_{353}$ map (as well as the other dust tracers). This suggests that the AME emissivity may not be a linear function of the dust column density in high-latitude cirrus regions---we discuss this more in \autoref{sec:discussion_a}.

We find that the AME emissivity, for most regions, is typically within 1 or 2\,$\sigma$ of the region averaged emissivity. There appears to be a bias where regions with larger uncertainties have larger emissivities (this is apparent in the skewed tail of the distribution shown in \autoref{fig:dust1_i}). Most of these regions are associated with latitudes of $b>50^{\circ}$, which is perhaps evidence that regions with lower total dust columns have a higher AME emissivity.


Finally, we discuss the measurements of dust-correlated emission measured at 4.76\,GHz in the \mbox{C-BASS} map using the $\tau_{353}$ dust template. We find that the weighted average of the dust coefficients measured in each region is consistent with zero ($\left<a_{4.76}\right>_\mathrm{regions} = (3 \pm 2)$\,K/$\tau_{353}$). We find some regions around the Galactic North pole do have a significant ($>3\sigma$) detection, but these regions have very little total emission in both the \mbox{C-BASS} and dust maps, and therefore we consider these to be unreliable estimates. When fitting all regions simultaneously we do make a detection of dust-correlated emission in the \mbox{C-BASS} map ($a_{4.76,\mathrm{all}} = (61 \pm 3)$\,K/$\tau_{353}$). When fitting for the dust at 4.76\,GHz, we found that the resulting free-free coefficients were 30\,per\,cent lower than when not fitting for the dust emission. There was no significant effect  on the synchrotron coefficients. The decrease in the free-free coefficients is likely due to the correlation between dust and the warm interstellar medium at high latitudes \citep{Dobler2008}. We therefore attempted fitting for the all-region dust coefficients after first subtracting an estimate of the free-free emission at 4.76\,GHz (assuming $T_e = 5000$\,K), we found that this made no significant change to the estimate of the AME emissivity in the \mbox{C-BASS} map. The AME emissivity at 4.76\,GHz relative to 22.8\,GHz is approximately what would be expected from the AME spectrum, however there are a large number of potential biases on such large-scales that could  account for the detection (e.g., residual free-free emission, dust-synchrotron correlation, etc.), so we consider this estimate to be an upper limit.

\subsection{Synchrotron}\label{sec:sync}

In the frequency range 0.408--22.8\,GHz the spectrum of Galactic synchrotron emission is expected to steepen with frequency due to CRE aging \citep[e.g.,][]{Strong2011}, an effect that has been observed in diffuse Galactic synchrotron emission in a number of observations \citep[e.g.,][]{Platania1998,Kogut2011}. In this section we will assess the curvature of the synchrotron spectrum by modelling the synchrotron spectrum as a simple power-law and looking at the change in the synchrotron spectral index when using either 408\,MHz or 4.76\,GHz data to trace the Galactic synchrotron emission.

We model the synchrotron spectrum as a power-law
\begin{equation}\label{eqn:sync1}
    T(\nu) = A_r \left(\frac{\nu}{\nu_r}\right)^{\beta_s},
\end{equation}
where $\beta_s$ is the spectral index of synchrotron spectrum between a given frequency $\nu$ and a reference frequency $\nu_r$, and $A_r$ is the synchrotron amplitude at $\nu_r$. The fitted synchrotron coefficients give a measure of the brightness ratio of the synchrotron emission between the template map (e.g., the 4.76\,GHz or 408\,MHz maps) and the map to be fitted; the coefficient is therefore defined as
\begin{equation}\label{eqn:sync2}
  a = \frac{T(\nu)}{T(\nu_0)} = \left(\frac{\nu}{\nu_0}\right)^{\beta_s},
\end{equation}
where $\nu_0$ is the frequency of the template map. The spectral index can then be found by simply rearranging \autoref{eqn:sync2}:
\begin{equation}\label{eqn:sync3}
    \beta_s = \frac{\log(a)}{\log(\nu/\nu_0)}.
\end{equation}
There are two systematic biases we have mitigated for when deriving spectral indices using \autoref{eqn:sync3}; details of these biases are given in \autoref{sec:sync_bias}. \autoref{tab:sync1} gives a summary of the spectral indices between the 408\,MHz/4.76\,GHz synchrotron templates and 22.8, 28.4, and 33\,GHz. A full summary of each regions spectral index measured at 22.8\,GHz can be found in \autoref{tab:all_coeff1}.

\begin{figure}
    \centering
    \includegraphics[width=0.45\textwidth]{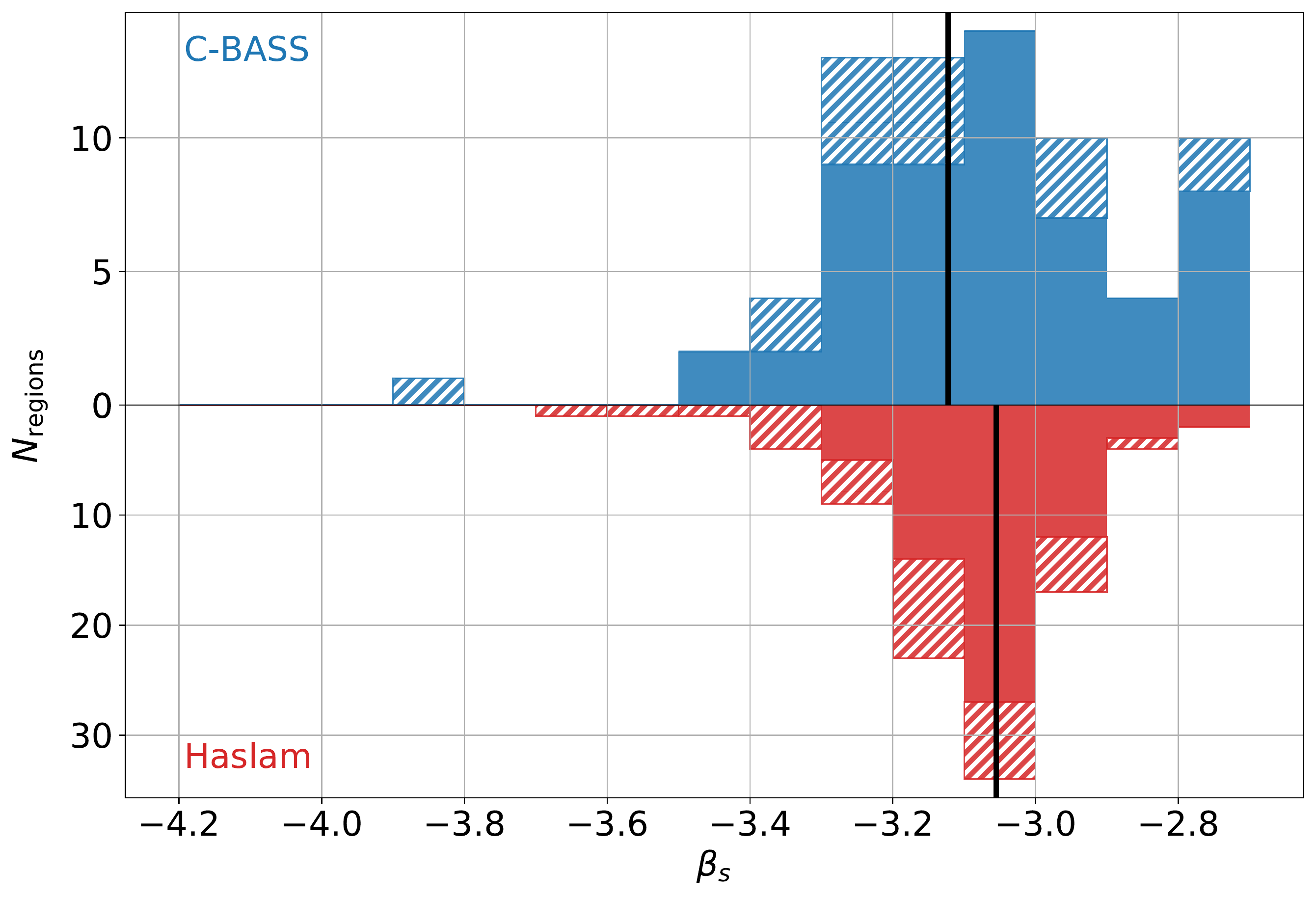}
    \caption{Distributions of spectral indices derived from the 22.8\,GHz WMAP K-band data with the 4.76\,GHz \mbox{C-BASS} and 408\,MHz Haslam datasets. The solid lines indicate the spectral index derived from the weighted mean of all the template coefficients. \textit{Hatched} regions indicate spectral indices with signal-to-noise ratios less than 5, and are not used to estimate the mean.}
    \label{fig:sync1}
\end{figure}

In \autoref{fig:sync1} we show the distributions of synchrotron spectral indices derived using \autoref{eqn:sync3} at 22.8\,GHz. The solid regions in the figure are spectral indices for regions where the synchrotron template coefficient derived using the 4.76\,GHz template has a signal-to-noise ratio greater than five. The difference in the mean spectral indices (the \textit{solid-black} lines) is $\Delta \beta_{22.8} = -0.06\pm0.02$ (\autoref{tab:sync1}). The steeper spectral indices measured at 4.76\,GHz suggest that, on average, Galactic synchrotron emission is steepening at mid-to-high Galactic latitudes with increasing frequency. This supports previous estimates of Galactic synchrotron curvature from historical surveys \citep{Platania1998}, COSMOSOMAS \citep{Fernandez2006}, and ARCADE2 \citep{Kogut2011}; there is also some evidence that Galactic synchrotron emission is flattening with increasing frequency in some regions of the Galaxy \citep{Peel2012}. We measure the average synchrotron spectral index between 4.76\,GHz and 22.8\,GHz to be $\left<\beta_{22.8}\right> = -3.10 \pm 0.02$, which is more consistent with the spectral index of polarized Galactic synchrotron emission measured between 22.8--100\,GHz  \citep[$\beta_s\approx-3.1$;][]{Dunkley2009,Fuskeland2014,Fuskeland2021} than that measured in intensity using the 408\,MHz data \citep[$\beta_s \approx -3.0$;][]{Banday2003,Davies2006,Ghosh2012}.

\begin{table}
\centering%
\renewcommand{\arraystretch}{1.5}
\caption{Region averaged and all-region spectral indices between 408\,MHz (\textit{top}) and 4.76\,GHz (\textit{bottom}) and 22.8, 28.4, and 33\,GHz datasets. The third and first quartiles of the spectral index distributions are given as super- and subscripts respectively. Uncertainties include both statistical and calibration uncertainties.}\label{tab:sync1}%
\begin{tabular}{l|c|c}%
\hline%
\hline%
&$\left< \beta \right>_\mathrm{regions}$&$\beta_\mathrm{all}$\\%
\hline%
$0.408 - 4.76$\,GHz&$-3.04^{-2.92}_{-3.16}\pm0.02$&$-3.02\pm0.02$\\%
$0.408 - 22.8$\,GHz&$-3.04_{-3.14}^{-2.98}\pm0.01$&$-3.02\pm0.01$\\%
$0.408 - 28.4$\,GHz&$-3.04_{-3.17}^{-2.98}\pm0.01$&$-3.01\pm0.01$\\%
$0.408 - 33$  \,GHz&$-3.04_{-3.18}^{-2.98}\pm0.01$&$-3.06\pm0.01$\\%
\hline%
$4.76 - 0.408$\,GHz&$-2.93_{-3.16}^{-2.90}\pm0.04$&$-3.01\pm0.04$\\%
$4.76 - 22.8$ \,GHz&$-3.10_{-3.32}^{-2.99}\pm0.02$&$-2.92\pm0.02$\\%
$4.76 - 28.4$ \,GHz&$-3.11_{-3.29}^{-3.02}\pm0.02$&$-2.91\pm0.02$\\%
$4.76 - 33$   \,GHz&$-3.09_{-3.29}^{-2.98}\pm0.02$&$-3.01\pm0.02$\\%
\hline%
\hline%
\end{tabular}%
\end{table}

\begin{figure*}
    \centering
    \includegraphics[width=0.6\textwidth]{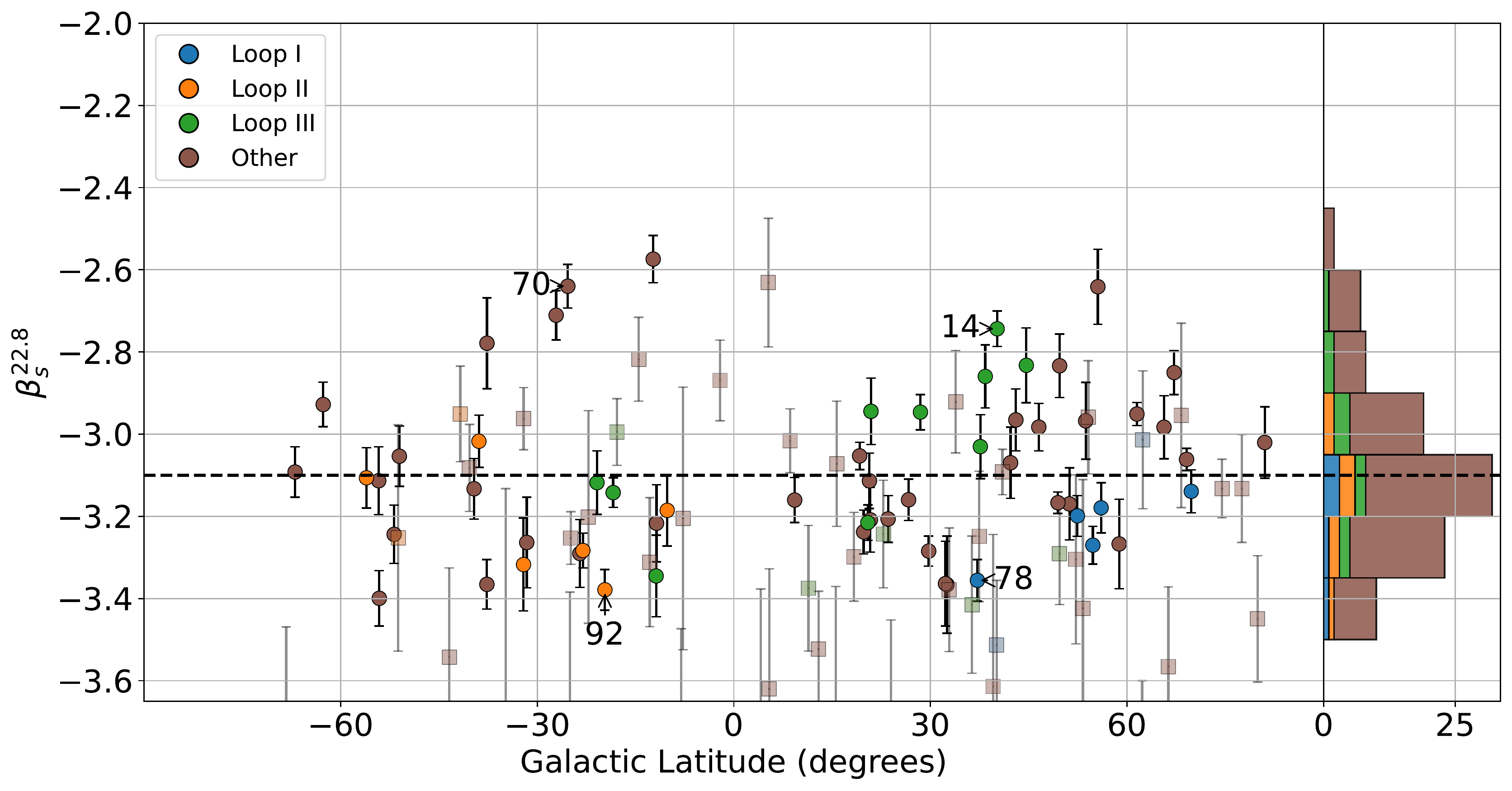}
    \includegraphics[width=0.6\textwidth]{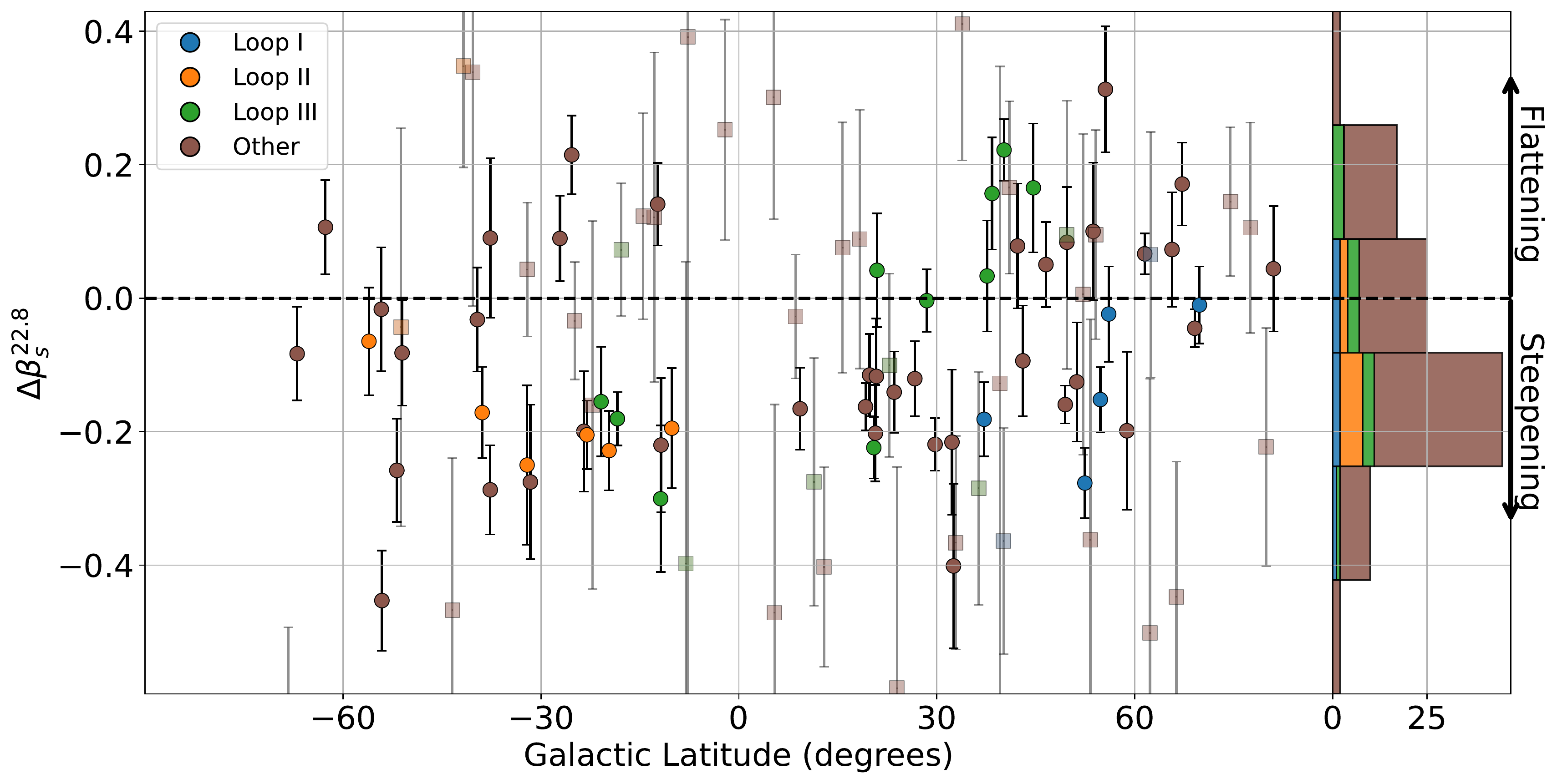}
    \caption{\textit{Top}: The spectral index in each region as measured between 4.76 and 22.8\,GHz. \textit{Bottom}: The change in the spectral index at 22.8\,GHz when using the 4.76\,GHz data instead of the 408\,MHz data to trace synchrotron emission, for each region. Colours indicate region groups associated with known Galactic synchrotron loops as defined in \autoref{fig:regions1}. Semi-transparent symbols represent regions that have a spectral index with a signal-to-noise of less than 5 when using either the 408\,MHz or 4.76\,GHz data. The \textit{black-dashed} line in the \textit{top} panel gives the weighted mean over all regions. The \textit{black-dashed} line in the \textit{bottom} panel indicates $\Delta \beta = 0$. Annotations indicate region numbers referenced in the main text.} 
    \label{fig:sync3}
\end{figure*}

In \autoref{fig:sync3} we show both the measured absolute spectral index between 4.76\,GHz and 22.8\,GHz (\textit{top}) and the change in the spectral index at 22.8\,GHz when using the 4.76\,GHz data instead of the 408\,MHz data to trace synchrotron emission (\textit{bottom}). We use colours in the figure to indicate region groups that are associated with known synchrotron loops in the northern hemisphere using the classification described in \citet{Vidal2015}. \autoref{fig:regions1} shows how these region groups are defined. The \textit{black-dashed} line in the \textit{top} panel of the figure shows the weighted mean over all regions.

In the \textit{top} panel of \autoref{fig:sync3} we can see there are several regions with extremely steep spectra---specifically the low latitude Loop\,I and Loop\,II regions with spectral indices of $\beta_\mathrm{LoopI} = -3.36\pm0.05$ (Region--78) and $\beta_\mathrm{LoopII}= -3.38\pm0.05$ (Region--92), respectively. The steep spectral index at the base of Loop\,I is especially striking as it implies that either the synchrotron emission within Loop\,I is flattening at high latitudes with increasing frequency (implying that the CREs at high latitudes are being reaccelerated) or we are measuring the spectral index from a superposition of different components along the line-of-sight at low Galactic latitudes; an argument that has been cited previously as a way of resolving discrepancies in the distance measurements to Loop\,I \citep{Panopoulou2021}. This is especially interesting since the spectral indices we measure using the 408\,MHz data around Loop\,I agree with those in the literature \citep[e.g.,][]{Davies2006}, implying it is not simply a product of region selection.

\autoref{fig:sync3} also shows a number of regions where the synchrotron spectrum hardens/flattens at high frequencies. The regions in the southern Galactic hemisphere that show significant flattening ($\sigma > 3$) are associated with bright H\textsc{ii} regions such as the Eridanus bubble (Region--70), implying the flattening is due to residual free-free emission in the 4.76\,GHz template in these regions.  In the northern Galactic hemisphere Region--14 shows a pronounced flattening of the spectrum, and is associated with synchrotron \mbox{Loop\,III}. This, like in the case of \mbox{Loop\,I}, is likely due to a superposition of components along the line of sight in this region, but in this instance it has resulted in a flattening of the synchrotron spectrum.  

Several of our regions coincide with the edge of the WMAP haze (Regions: 91, 105, 106), a region of diffuse hard spectrum emission ($\beta \approx -2.5$) in total intensity coincident with the Galactic centre \citep{Finkbeiner2004b,Dobler2008,planck2012-IX}. We find that the average fitted synchrotron spectral index in these regions is consistent with the average for the rest of the sky ($\left<\beta_s\right>_\mathrm{Haze} = -3.11 \pm 0.05$). Likewise, we find that these regions show no significant background residuals (i.e., there is no evidence that the haze component is not being fitted). A full exploration of the haze emission at 4.76\,GHz will require the upcoming southern \mbox{C-BASS} survey.

\subsection{Free-Free}\label{sec:halpha}

At mid-to-high latitudes it is challenging to constrain diffuse free-free emission as it is extremely faint except in several nearby star forming regions such as Orion and Eridanus. Free-free emission in the \mbox{C-BASS} map is approximately 25$\times$ brighter than in the WMAP K-band map in units of brightness temperature. Even so, for most regions analysed here, we found that the H$\alpha$-related coefficients have low S/N due to the free-free emission being a relatively small component relative to synchrotron/AME. The only exceptions were bright ionized regions such as those associated with Orion or the Eridanus loop. Therefore in order to maximize the signal-to-noise of the free-free emission coefficient at 4.76\,GHz we present just the results for the all-region fit only. We do this for two cases, with and without a correction for dust absorption.

The relationship between the free-free radio brightness and H$\alpha$ intensity is well defined in regions that have little H$\alpha$ line-of-sight dust absorption or H$\alpha$ scattering, and a known  electron temperature. The relationship is defined as \citep{Draine2011}
\begin{equation}\label{eqn:freefree1}
    \frac{T_b^{\rm ff}}{I_{\mathrm{H}\alpha}} = 1512\, T_4^{0.517} 10^{0.029/T_4} \nu_\mathrm{GHz}^{-2} g_{\rm ff},
\end{equation}
where $T_4$ is the electron temperature ($T_e$) in units $10^4$\,K, and $g_{\rm ff}$ is the Gaunt factor. The Gaunt factor accounts for the quantum mechanical effects and is approximated by
\begin{equation}
    g_{\rm ff} = \mathrm{ln}\left(\mathrm{exp}\left[5.960 - \frac{\sqrt{3}}{\pi}\mathrm{ln}\left(\nu_\mathrm{GHz} T_4^{-3/2} \right) \right] + e \right).
\end{equation}

Fitting for the free-free emission in the 4.76\,GHz map using the 408\,MHz map to trace synchrotron emission and the H$\alpha$ map with no correction for dust absorption we find a free-free to H$\alpha$ ratio of $T_b^{\rm ff}/I_{\mathrm{H}\alpha} = (258 \pm 5)$\,$\mu$K/R\footnote{We also fitted the data using the H$\alpha$ map provided on the NASA LAMBDA website by \citet{Finkbeiner2003} and find a similar ratio of $T_b^{\rm ff}/I_{\mathrm{H}\alpha} = (241 \pm 9)$\,$\mu$K/R.}, which corresponds to an electron temperature of $T_e = (5600 \pm 130)$\,K. If we make a correction for dust absorption assuming that the dust and gas are mixed \citep[e.g., $f_d=0.33$;][]{Dickinson2003} we find a ratio of $T_b^{\rm ff}/I_{\mathrm{H}\alpha} = (195 \pm 5)$\,$\mu$K/R which corresponds to $T_e = (3300 \pm 120)$\,K. In both cases the global average electron temperature is lower than the $6000 < T_e < 8000$\,K range that is expected from radio-recombination lines \citep[e.g.,][]{Alves2012,Alves2015} or optical line ratios \citep[e.g.,][]{Haffner1999}. 

Lower-than-expected estimates of the temperature of the WIM from comparison between radio and H$\alpha$ observations have been measured many times before \citep{Banday2003,Davies2006,Dobler2008,planck2014-a31}. There have been several attempts to explain the discrepancy between measured electron temperature of the WIM using continuum free-free emission, and those measured with optical line ratios or RRLs via the physical conditions of the WIM \citep{Dong2011,Geyer2018}, however the current most favoured explanation is that between 10 to 50\,per\,cent of the observed H$\alpha$ intensity actually comes from secondary scattering from nearby dust clouds \citep{Wood1999,Witt2010,Brandt2012,Barnes2015,planck2014-a31}. We estimate the scattering fraction by comparing the mean measured H$\alpha$ coefficients above with the theoretical estimate for H$\alpha$ emission assuming an electron temperature range of $6000 < T_e < 8000$\,K, which corresponds to a range of free-free to H$\alpha$ ratios of $270 < T_b^{\rm ff}/I_{\mathrm{H}\alpha} < 320$\,$\mu$K/R at 4.76\,GHz. We can then calculate the fraction of scattered H$\alpha$ light as
\begin{equation}
    f_{\mathrm{sc}} = \frac{a_0}{\hat{a}} - 1,
\end{equation}
where $\hat{a}$ is the measured free-free brightness to H$\alpha$ intensity and $a_0$ is the theoretical value given by \autoref{eqn:freefree1}. We find that if we assume no dust absorption that the fraction of scattered light has a range of  $0.04 < f_{\mathrm{sc}} < 0.23$, while after correcting for the dust absorption assuming the dust and gas are mixed we get the range $0.38 < f_{\mathrm{sc}} < 0.64$.

As mentioned before, most individual regions do not have a significant detection of the free-free to H$\alpha$ ratio, except in the Orion-Eridanus regions. If we exclude the Orion-Eridanus regions, the weighted mean of the free-free coefficients for the remaining regions is still less than expected $T_b^{\rm ff}/I_{\mathrm{H}\alpha} = (266 \pm 6)$\,$\mu$K/R or a $T_e = (5900\pm160)$\,K (assuming no dust absorption). Similarly for just the Orion-Eridanus region we find the ratio is also low $T_b^{\rm ff}/I_{\mathrm{H}\alpha} = (243 \pm 9)$\,$\mu$K/R or $T_e = (5000\pm230)$\,K. We can therefore see that the underestimate of the free-free to H$\alpha$ ratio seems to affect all of the mid-to-high latitude sky observable by \mbox{C-BASS}. This smaller than expected ratio implies that either the WIM is, in most instances, in front of the dust along most lines of sight and not being absorbed or the fraction of scattered H$\alpha$ light is significantly higher than previously estimated using WMAP/\textit{Planck} data \citep{planck2014-a31}. 


\vspace{-0.1cm}
\section{Dust Model Fitting}\label{sec:fitting}

The fitted dust template coefficients are dependent on the relative brightness of the radio emission relative to the dust template. As such the fitted coefficients are encoded with the relative spectrum of the underlying dust grain population. There are two main components to the spectrum of the fitted dust coefficients: At lower frequencies ($\nu < 100$\,GHz) the dust coefficient spectrum is dominated by contributions from spinning dust emission, while at higher frequencies the spectrum is dominated by contributions from thermal dust emission. With WMAP and \textit{Planck} data alone it is not possible to constrain the peak frequency of the AME spectrum at high latitudes \citep[e.g.,][]{planck2014-a12}, we therefore use dust template fits to the \mbox{C-BASS} map (fitted using the 408\,MHz data to remove the synchrotron component---\autoref{sec:dust}) to constrain the low frequency turn over in the AME spectrum. 

We fit the dust coefficients discussed in \autoref{sec:dust} with a two component model in brightness temperature units defined as
\begin{equation}\label{eqn:dmodel1}
    a(\nu) = a_\mathrm{td}(\nu) + a_\mathrm{sp}(\nu),
\end{equation}
where $a_\mathrm{td}(\nu)$ is the contribution to the dust coefficient from thermal dust emission, and $a_\mathrm{sp}(\nu)$ is the contribution from spinning dust. We model the thermal dust component as a modified blackbody curve normalized at 353\,GHz as
\begin{equation}
    a_\mathrm{td}(\nu) = A_\mathrm{td} \left(\frac{\nu}{\nu_0}\right)^{\beta_\mathrm{d} + 1} \frac{\exp(\gamma \nu_0)-1}{\exp(\gamma \nu)-1},
\end{equation}
where $A_\mathrm{td}$ is the value of the dust coefficient at 353\,GHz, \mbox{$\gamma = h/k_B T_\mathrm{d}$} (where $T_\mathrm{d}$ is the dust temperature) and $\beta_\mathrm{d}$ is the thermal dust spectral index. We place Gaussian priors on the thermal dust temperature ($T_\mathrm{d}$) and $\beta_\mathrm{d}$ parameters of $N(20.3 \pm 5\,\mathrm{K})$ and $N(1.59 \pm 0.5)$, respectively, as these parameters have been well constrained by previous analyses \citep[e.g.,][]{planck2013-p06b,planck2014-a12}.

We use two models for the AME spectrum. The first jointly fits two spinning dust spectra generated using the \texttt{SpDust2} model  \citep{AliHaimoud2009,Silsbee2011} assuming two different environmental conditions: the cold neutral medium (CNM) and the warm neutral medium (WNM). The typical dust parameters for the CNM and WNM phases were taken from \mbox{table~1} of \citet[][]{Draine1998a}. The approach of combining two \texttt{SpDust2} spectra to model multiple grain populations along a line-of-sight has been shown to be effective at modelling the AME spectrum in the past \citep[e.g.,][]{Ysard2010,planck2013-XV}. However, to allow for more flexibility in the model we allow the \texttt{SpDust2} spectra to be interpolated to different frequency ranges, thus allowing for us to fit for a peak in the spinning dust spectrum. The spinning dust model can be parametrized as
\begin{equation}\label{eqn:dmodel}
    a_\mathrm{sd}(\nu) =  \left(\frac{\nu_0}{\nu}\right)^2   \frac{\sum_{i} A_i f_i(\nu / \nu_{pi}\nu_{p0})}{\sum_{i} f_i(\nu_0 / \nu_{pi}\nu_{p0})},
\end{equation}
where we are summing over the CNM and WNM models,  $A_i$ is the amplitude of each spectrum at the reference frequency of \mbox{$\nu_0=22.8$\,GHz}, $f_i$ is the interpolated \texttt{SpDust2} spectrum functions, and $\nu_{pi}$ is the peak frequency of each spectrum. The $\nu_{p0}$ parameter is the peak frequency of the input  \texttt{SpDust2} CNM and WNM spectra. 

For the \texttt{SpDust2} model we must enforce several priors since we still have a limited set of data for constraining the AME spectrum. We follow a similar approach as outlined in \citet{planck2014-a12}, but allow for looser priors due to the additional constraints given by \mbox{C-BASS} at low frequencies. First, we set the condition that the peak frequency of the CNM component is less than the peak of the WNM component ($\nu_\mathrm{CNM} < \nu_\mathrm{WNM}$), further we enforce positivity on all components. For the peak frequency of the WNM component we set a weak Gaussian prior of $N(30\pm15 \mathrm{GHz})$ to ensure that the WNM component is not biased to fit the thermal dust emission. For the peak frequency of the CNM we enforce no prior, allowing for the data to drive the determination of the peak frequency.\footnote{Fits are performed using the Affine Invariant Markov chain Monte Carlo Ensemble sampler \textsc{emcee} \citep{Foreman-Mackey2013}.}

For the second second model we use a two parameter log-normal distribution, which has been shown to well characterize the AME spectrum \citep{Bonaldi2007,Stevenson2014,Cepeda-Arroita2021}. Here we use the model first introduced by \citet{Bonaldi2007} of
\begin{equation}\label{eqn:bonaldi}
   \begin{aligned}
  \log(a_\mathrm{sd}) =  -\left(\frac{m_{60} \log (\nu_\mathrm{p})}{\log(\nu_\mathrm{p}/60~\mathrm{GHz})}\right) \log\left(\frac{\nu}{23~\mathrm{GHz}}\right) \\
 + \frac{m_{60}}{2 \log(\nu_\mathrm{p}/60~\mathrm{GHz})}\left[\log(\nu/1~\mathrm{GHz})^2 - \log(23/\mathrm{GHz})^2\right] \\
 + 2\log\left(\nu / 23~\mathrm{GHz}\right) + \log(A)
   \end{aligned}
\end{equation}
where $\nu_p$ is the peak frequency of the spectrum, $A$ is the amplitude at 23\,GHz and $m_{60}$ is the spectral index at 60\,GHz. As with the \texttt{SpDust2} model we enforce positivity for all parameters, we also must enforce that the peak frequency is less than 60\,GHz since \autoref{eqn:bonaldi} is not defined at 60\,GHz. We use no Gaussian priors on the parameters. 

\begin{figure}
    \centering
    \includegraphics[width=0.4\textwidth]{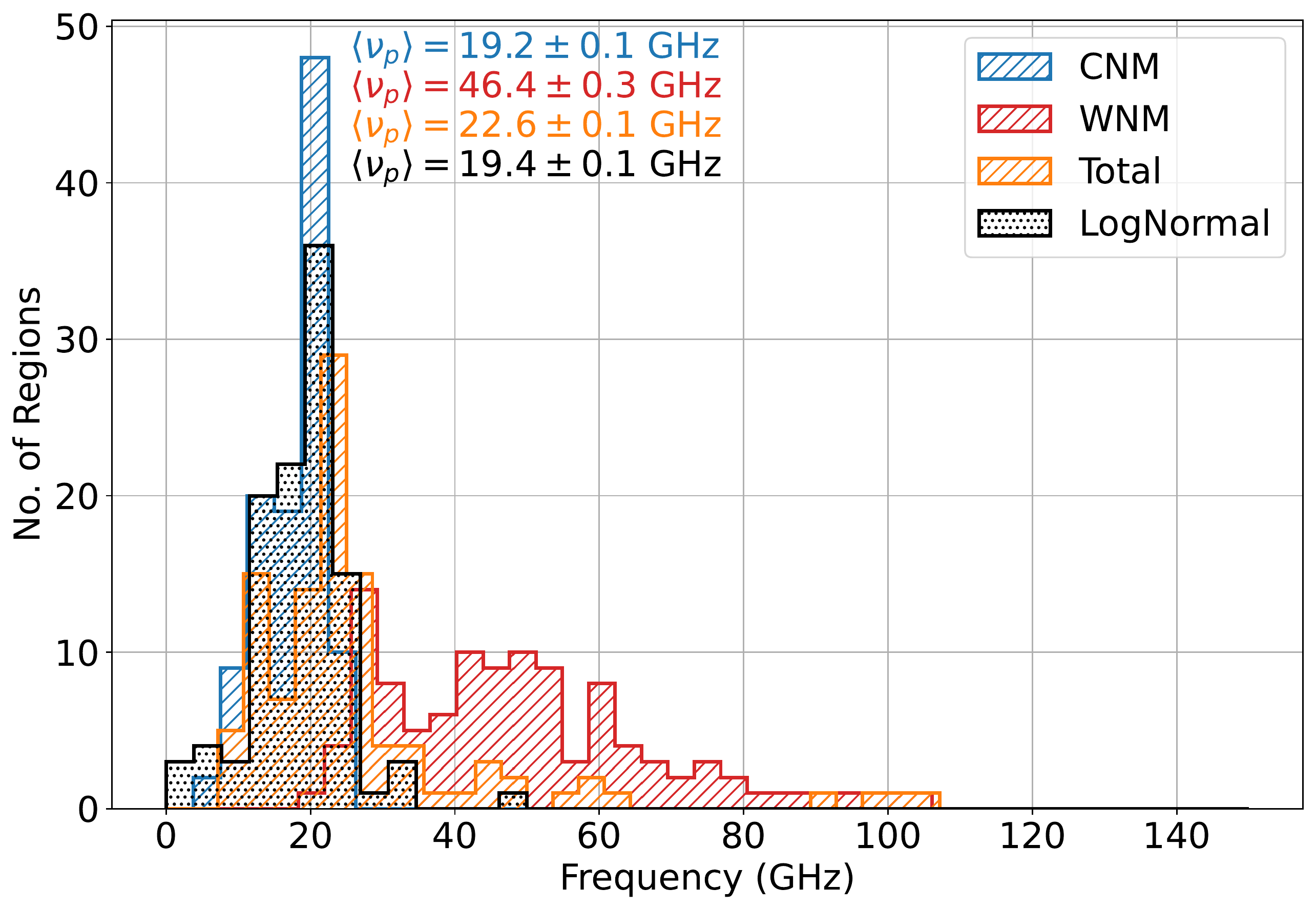}
    \caption{Flux density peak frequencies of the individual CNM and WNM components, and for the combined spinning dust spectra.}
    \label{fig:peak_freq_dmodels}
\end{figure}


\begin{table}%
\caption{Best-fitting parameters for the log-normal and two component \texttt{SpDust2} models. The first uncertainty represents the standard deviation of each distribution shown in \autoref{fig:peak_freq_dmodels} and the second is the uncertainty in the distribution mean.}%
\label{tab:dustmodel1}%
\centering%
\begin{tabular}{lclc}%
\hline%
\hline%
Parameters&\mbox{log-normal}&Parameters&\texttt{SpDust2}\\%
\hline%
$\nu_\mathrm{p}$ [GHz]&$ 19.4 \pm 6.4 \pm 0.1$&$\nu_\mathrm{CNM}$ [GHz]&$ 19.2 \pm 4.5 \pm 0.1$\\%
$m_{60}$&$ 3.5 \pm 1.7 \pm 0.1$&$\nu_\mathrm{WNM}$ [GHz]&$ 46.4 \pm 18.5 \pm 0.3$\\%
&&$\nu_\mathrm{p}$ [GHz]&$ 22.6 \pm 17.7 \pm 0.1$\\%
\hline%
\end{tabular}%
\end{table}

In \autoref{fig:peak_freq_dmodels} we show the best-fitting peak frequencies for both the \texttt{SpDust2} and log-normal models in flux density units \footnote{The spinning dust spectrum is not peaked in brightness temperature units.}. For the \texttt{SpDust2} model we also show the fitted CNM and WMN components peak frequencies. We find that the median peak frequency of the AME flux density spectrum for the \texttt{SpDust2} and log-normal models agree on average at $(19.4 \pm 0.1)$\,GHz and $(22.6 \pm 0.1)$\,GHz respectively (see \autoref{tab:dustmodel1}). Similar peak frequencies for high latitudes regions ($|b| > 10^\circ$) were found by the \textit{Planck} \texttt{Commander} analysis \citep{planck2014-a31}, but the Gould belt region was found to have a higher mean peak frequency of $(25.5\pm1.5)$\,GHz \citep{planck2013-XII}. Comparing the reduced chi-squared ($\chi_r^2$) of each model shows no strong preference for the more complex three-parameter \texttt{SpDust2} model ($\chi_r^2 = 2.4$ ) over the two-parameter log-normal model ($\chi_r^2=2.9$). 


\begin{figure}
    \centering
    \includegraphics[width=0.4\textwidth]{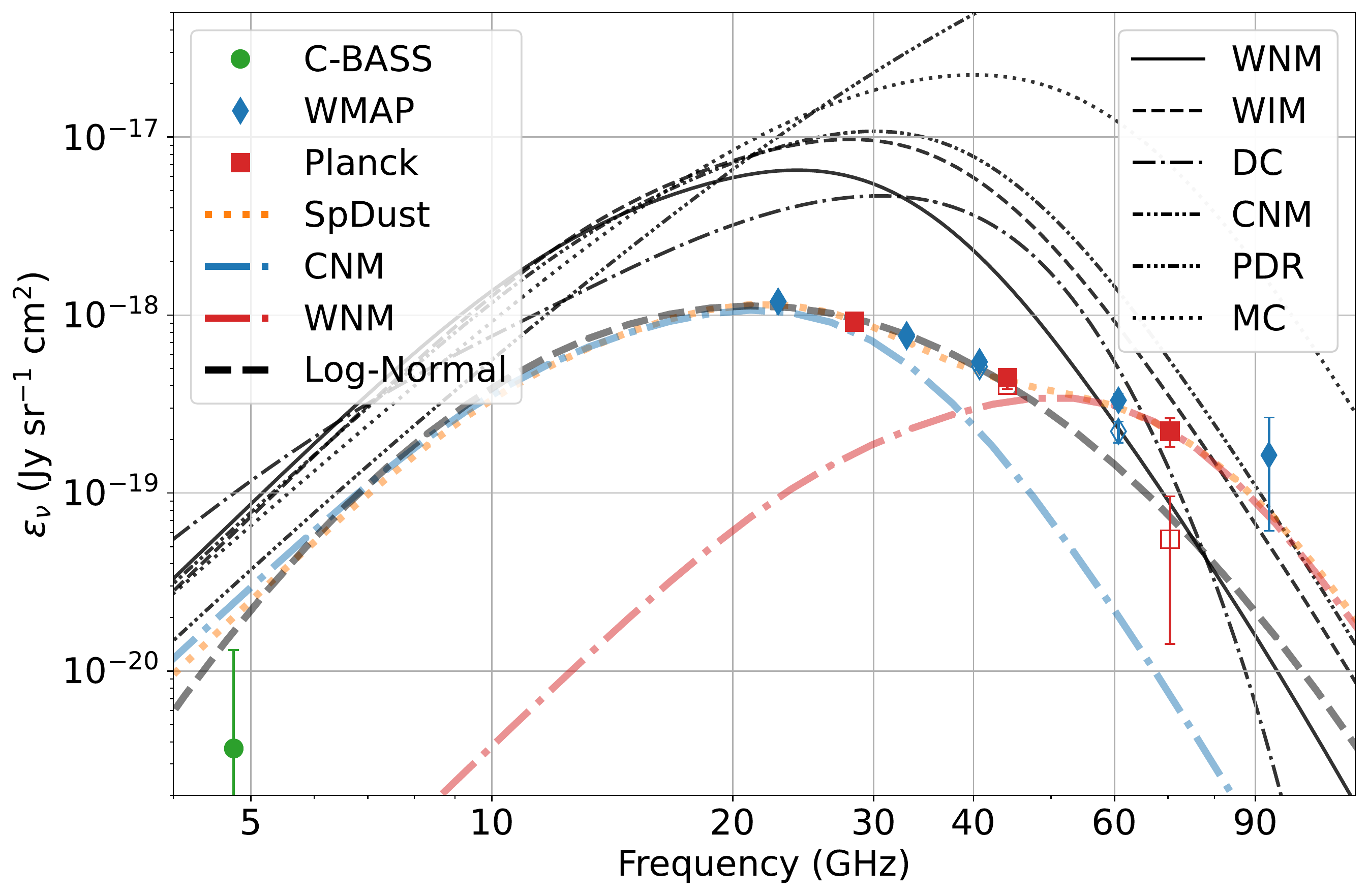}
    \caption{AME emissivity spectrum showing the best-fitting \texttt{SpDust2} and log-normal models to the weighted average dust coefficients over all regions. The 4.76\,GHz \mbox{C-BASS} data are represented by the \textit{green} circles, the \textit{blue} diamonds are WMAP, and the \textit{red} squares are \textit{Planck}. Solid markers are associated with \texttt{SpDust2} model fit, and hollow markers with the log-normal fit. Six \texttt{SpDust2} spinning dust spectra are shown for generic environments: warm neutral medium (WNM), warm ionised medium (WIM), dark clouds (DC), cold neutral medium (CNM), photodissociation regions (PDR), and molecular clouds (MC). }
    \label{fig:dmodel1}
\end{figure}

In \autoref{fig:dmodel1} we show the best-fitting \texttt{SpDust2} and log-normal models fitted to the weighted average dust coefficients in units of emissivity (i.e., brightness per unit column density---see \autoref{eqn:dust_emiss})  with the estimate of the thermal dust contribution subtracted. The 4.76\,GHz \mbox{C-BASS} data are critical for constraining the AME spectrum in these plots as they constrain the low frequency turnover. Without \mbox{C-BASS} there is effectively no constraint on the AME peak frequency at high latitudes without using strong priors \citep{planck2014-a31}. These constraints could be further improved by including data from the QUIJOTE 10--20\,GHz northern sky survey \citep[e.g.,][]{Genova2015}.

\autoref{fig:dmodel1} shows our estimates of the AME emissivity using the $\tau_{353}$ template alongside \texttt{SpDust2} spinning dust spectra for six generic environments based on the parameters in \citet[Table 3,][]{Draine1998a}. We find that the measured emissivities are approximately an order-of-magnitude lower than the predictions from the \texttt{SpDust2} models. However, this discrepancy can be explained by a number of factors mostly related to the calculation of the dust column density required to estimate the AME emissivity. Taking into account all factors, the differences between the AME emissivity we measure and the predictions from the generic models can be explained. We will discuss this in more detail in \autoref{sec:discussion_a}.

\vspace{-0.1cm}
\section{Discussion}\label{sec:discussion}

We have shown how using the \mbox{C-BASS} 4.76\,GHz data in place of the 408\,MHz data to trace synchrotron emission in the WMAP and \textit{Planck} maps results in very little change in the dust emissivity measured around 30\,GHz. We have also shown how \mbox{C-BASS} data suggests that on average the synchrotron spectrum is steepening at high frequencies. In this section we will discuss these findings in the wider context of findings given in the literature.

\subsection{High Latitude Anomalous Microwave Emission}\label{sec:discussion_a}

One of the earliest detections of anomalous microwave emission was at high Galactic latitudes \citep{Kogut1996b} yet confirming the nature and origin of AME at high latitudes has proven to be more challenging than for individual compact objects due to the lack of all-sky surveys between 2\,GHz and WMAP/\textit{Planck} frequencies. As such it is somewhat more difficult to rule out alternative AME hypotheses such as a dust-correlated hard/flat spectrum synchrotron component that is associated with Galactic star formation \citep{Bennett2003}. A clear test to verify whether there is a flat spectrum synchrotron component that correlates with dust would be to observe such a feature in a higher frequency tracer of synchrotron. \citet{Peel2012} looked for evidence of a hard synchrotron component using 2.3\,GHz data but they found no significant evidence for its presence. We find a similar result using \mbox{C-BASS} (as discussed in \autoref{sec:dust}). Although there is a tentative detection of AME when fitting all regions simultaneously, it is found to have a rising spectrum between 4.76 and 22.8\,GHz which is inconsistent with flat-spectrum synchrotron.

\begin{figure}
    \centering
    \includegraphics[width=0.4\textwidth]{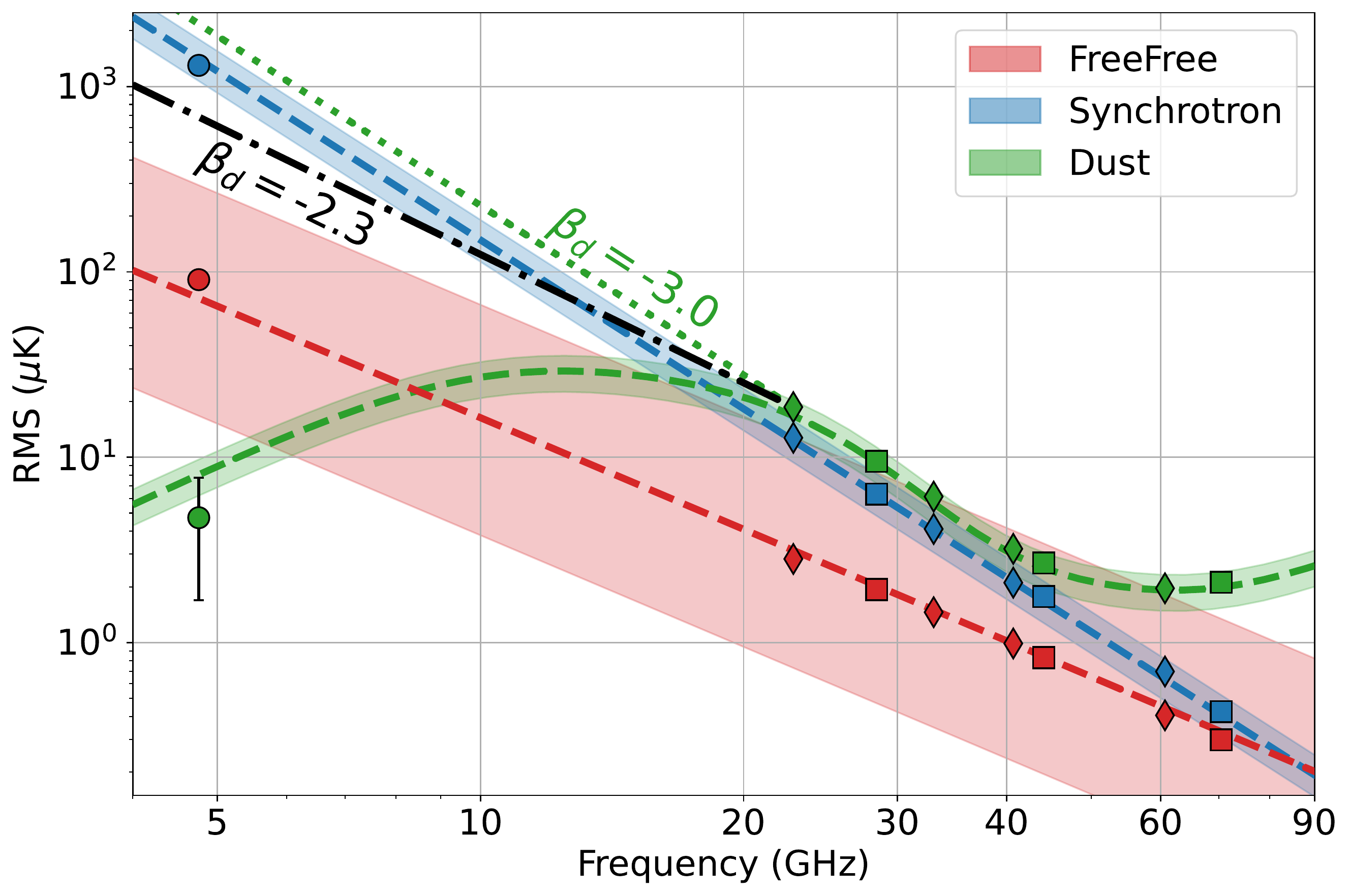}
    \caption{RMS fluctuations spectrum in Rayleigh-Jeans units for all foreground components at $1^{\circ}$ resolution. The symbols are the product of the weighted average coefficients and the template RMS with synchrotron as \textit{blue}, AME and thermal dust as \textit{green}, and free-free as \textit{red}. The \textit{green dashed} curve is  the best-fitting \texttt{SpDust2} plus thermal dust model, the \textit{blue dashed} curve is a power-law with spectral index $\beta_s=-3.0$ (from \autoref{tab:sync1}, and the \textit{red dashed} curve is the best-fitting power law free-free model. The shaded regions represent the region-to-region variations by scaling the models to the upper and lower percentiles of the region values at 22.8\,GHz for dust and synchrotron, and 4.76\,GHz for free-free. We also included the \textit{green dashed} line indicating the dust spectral index between 22.8 and 40\,GHz, and the \textit{black dash-dot} line to represent a flat synchrotron spectrum. \textit{Diamonds} are WMAP, \textit{squares} are Planck, and \textit{circles} are \mbox{C-BASS}.}
    \label{fig:discuss1}
\end{figure}

In \autoref{fig:discuss1} we show the spectrum of the region averaged coefficients for synchrotron, dust, and free-free emission scaled by the RMS of the associated templates in each region. In the figure we are using the $\tau_{353}$ template for the dust, and the 4.76\,GHz template for the synchrotron emission. By doing this we are able to directly compare the relative brightnesses of each emission component. The spectral index of the dust emission between 22.8\,GHz and 40\,GHz is $\beta \approx -3$ (shown in the figure as the green dashed line) shows that if the dust emission can be modelled by a power law it would overestimate the limits placed on the AME spectrum by more than two orders of magnitude. We also extrapolate the amplitude of the dust spectrum at 22.8\,GHz using a power law spectrum with $\beta = -2.3$ (black dot-dashed line) indicative of flat-spectrum synchrotron, but still find that this greatly exceeds the limits set by \mbox{C-BASS}. Given these results, high latitude AME cannot be explained by a flat-spectrum synchrotron component, and strongly favours the spinning dust hypothesis of AME.

Next, in \autoref{sec:dust} we found that the spinning dust component is traced equally well by all of the dust tracers that we used. This is somewhat of a contradiction to the findings of \citet{Hensley2016} where they find that spinning dust emission most strongly correlates with the \textit{Planck} derived radiance map. However, there are several differences between the two analyses;  \citet{Hensley2016} compared the 30\,GHz spinning dust map derived from the \texttt{Commander} analysis \citep{planck2014-a31} to the radiance map across all pixels simultaneously, while we are using the original WMAP and \textit{Planck} data in approximately 100\,deg$^{2}$ patches of the sky. Also interesting to note is that our $\tau_{353}$ coefficient is \textit{smaller} than they report, while our radiance coefficient is \textit{larger}. These discrepancies are hard to reconcile; however, at high latitudes the \texttt{Commander} spinning dust map is known to be limited due to difficulties in separating the spinning dust from diffuse free-free and synchrotron components, sometimes overestimating and sometimes underestimating the AME by approximately 30--40\,per\,cent \citep{planck2014-a31,Cepeda-Arroita2021}, which is of order the differences we see between our coefficients and those given in \citet{Hensley2016}. 

The environmental conditions of the high latitude regions we are looking at are known to be similar \citep[][]{planck2013-p06b}, possibly owing to the dust in all of these regions being located relatively nearby \citep[within approximately 0.5\,kpc,][]{Green2019}. The spinning dust spectrum is not expected to be very sensitive to the interstellar radiation field \citep{AliHaimoud2009,Ysard2010}, however in some AME sources the ISRF intensity has been found to correlate well with AME intensity \citep{Tibbs2011,Tibbs2012}.  \citet{planck2013-XV}  found that there is a limited correlation between AME emissivity, and AME peak frequency with the ISRF, which may be due to the emission from very small grains (VSGs) and PAHs being proportional to the ISRF strength \citep{Sellgren1985}. The relative strength of the ISRF can be parametrized by $G_0 = (T_D/17.5~\mathrm{K})^{4+\beta_D}$ \citep{Mathis1983}, the average value of $G_0$ for Galactic latitudes of $|b| > 15^{\circ}$ is $G_0 = 2.3$ \citep[Table 3;][]{planck2013-p06b}, which is higher than the typical $G_0$ found in denser, discrete sources nearer the Galactic plane \citep{planck2013-XV}. This implies that the high latitude dust clouds, though they may share a common environment with each other, is atypical of the environments associated with other sources of AME. One particularly interesting exception is the $\rho$~Ophiuchi molecular cloud that has a similar ISRF to high latitude clouds \citep{planck2013-XV}, and has a broad AME spectrum similar to \autoref{fig:dmodel1} \citep[e.g.,][]{planck2011-7.2}. The AME in $\rho$~Ophiuchi has been found to be associated with the PDR along the western edge of the cloud \citep{Casassus2008,Casassus2021}, which may indicate that the AME in high latitude dust clouds originates from similar PDR-like environments.

The measured AME emissivities shown in \autoref{fig:dmodel1} were found to be an order-of-magnitude lower than the generic environment spinning dust models (the \textit{black} lines in the figure). On the other hand, earlier analyses found a much better agreement with the models to within a factor of a few \citep[e.g.,][]{Draine1998b,Finkbeiner2004b}, which can be easily accounted for by adjusting the environmental parameters. Nevertheless, much of the difference in emissivities with previous analyses can be accounted for by our use of $\tau_{353}$ as a dust tracer. Most earlier analyses used the 100\,$\mu$m IRAS map as a tracer and a column density calibration of $0.85 \times 10^{-20}$\,MJy\,sr$^{-1}$\,cm$^{2}$ \citep{Boulanger1988}, which we find results in  differences in the column density with $\tau_{353}$ \citep{planck2013-p06b} by a factor of $\approx 2$. Additionally, early analyses typically fitted for the full sky as opposed to smaller regions, and indeed we find that our all-region fit for $I_{100\mu\mathrm{m}}$ is higher than the region average by approximately 50\,per\,cent. Together these effects bring our results into approximate agreement with previous results.

Nevertheless, our AME emissivities are still lower than the predictions of the spinning dust model. We first note that the spinning dust models are for generic interstellar environments using the parameters suggested in \citet{Draine1998a}, which need to be tuned to fit the observations. An order-of-magnitude can easily be achieved as demonstrated by \citet{Vidal2020}. Another important consideration is the averaging of the dust column (and also other quantities) over the regions, which may not be easily characterized by a single  \texttt{SpDust2} model parametrization. For example, if most of the AME originates on small scales, but on large scales there is none, then we would be underestimating the emitting dust column. To quantify this effect will require a multi-scale analysis of a known AME emitting region, which we will investigate in a future work.

\subsection{Synchrotron Curvature}

It has been known for some time that synchrotron spectra exhibit some degree of curvature. At frequencies between 100\,MHz and 1\,GHz the typical spectral index of synchrotron emission is  $\beta_s = -2.7$ \citep{Lawson1987,Platania1998,Platania2003}, while at frequencies between 408\,MHz and 23\,GHz the spectral index is  $-3.2  < \beta_s < -2.9$ \citep{Banday2003,Davies2006,Gold2011,Ghosh2012}. Previous attempts to measure the spectral index between microwave frequencies and  1--2\,GHz surveys have not found strong evidence for an increase in the curvature of the synchrotron spectrum \citep{Peel2012}. 

 The Galactic synchrotron spectrum tends to flatten at  frequencies below $< 10$\,GHz \citep[][]{Lawson1987,Platania1998,Rogers2008} in line with the expectations of CRE propagation models with a general steepening of the spectrum with increasing Galactic latitude indicative of CRE aging \citep{Orlando2013}. In \autoref{sec:sync} we find that the synchrotron spectrum between 4.76 and 22.8\,GHz is steeper than the spectrum between 0.408 and 22.8\,GHz by $\Delta \beta_s = -0.06 \pm0.02$ when averaged across all regions. For the all-region fits we generally find flatter spectral indices; we tested to see if this could be driven by the flatter spectral indices of the North Polar Spur by repeating the analysis with this region masked but found the spectral index with and without the North Polar Spur to be consistent. This difference is possibly due to systematic errors on large-scales or even real Galactic emission---we intend to investigate this further in the future.

\section{Conclusion}\label{sec:conclusion}

In this study of Galactic AME, synchrotron, and free-free emission using the \mbox{C-BASS} North data we have divided the sky into a \texttt{HEALPix} grid of $N_\mathrm{side}=4$ or regions of order 200\,deg$^2$ and applied the template fitting technique. 

The major result from this study is that we find the fitted dust coefficients around $\approx 30$\,GHz that trace AME do not significantly change ($<1$\,per\,cent on average) when using the \mbox{C-BASS} 4.76\,GHz map to trace the underlying synchrotron emission instead of the 408\,MHz synchrotron template. Further, we find that there is very little evidence for AME at 4.76\,GHz and place an upper limit on the emissivity of  $\epsilon_{4.76\mathrm{GHz}} < (2.9\pm0.2) \times 10^{-19}$\,Jy\,sr$^{-1}$\,cm$^{2}$\,H$^{-1}$ from an all-region fit. Together these two conclusions strongly disfavour AME at high Galactic latitudes being generated by  a hard/flat synchrotron component, and prefer instead that spinning dust is the origin of AME. We find that the fitted AME spectrum is well described by peaked spectrum models with a median peak frequency of $(22.6\pm0.1)$\,GHz (in flux density units)   when fitted using a two component \texttt{SpDust2} model and $(19.4\pm0.1)$\,GHz when fitted using a log-normal model, implying that the environment of high latitude cirrus clouds is different to those found in other sources of AME, which tend to peak around $\sim30$\,GHz.

We have found that when using the \mbox{C-BASS} 4.76\,GHz data to trace synchrotron on scales of $\approx\!10^{\circ}$ the synchrotron spectrum steepens at high frequencies by $\Delta \beta_s = -0.06\pm0.02$ when the reference frequency is changed from 4.76\,GHz to 408\,MHz. We also found that some regions such as Loop I exhibit substantial steepening at high frequencies, with the base of Loop I having an extreme spectral index of $\beta_s = -3.36 \pm 0.05$. However, in general the typical differences between the spectral indices at 22.8\,GHz derived using the 408\,MHz ($\left<\beta_s\right>_{408}=3.04 \pm 0.01$) or 4.76\,GHz \mbox{C-BASS} ($\left<\beta_s\right>_{4.76}=-3.1\pm0.02$) data are small, and we find no evidence for a hard spectrum synchrotron component (that would result in a flattening of the spectrum at high frequencies). Therefore extrapolation of the Galactic synchrotron to high frequencies remains a reasonable model, which is important for CMB foreground removal.

For free-free emission, we put new constraints on the free-free brightness to H$\alpha$ intensity using the 4.76\,GHz \mbox{C-BASS} data. Due to the faintness of the free-free emission at intermediate and high Galactic latitudes we were only able to get a significant detection on the free-free/H$\alpha$ ratio when performing a fit across all region simultaneously, or by selecting bright H\textsc{ii} structures within the Orion-Eridanus regions. We tested fitting for the free-free emission using H$\alpha$ templates with either no dust absorption correction and a dust absorption correction assuming a dust mixing fraction of $f_D = 0.33$ and find electron temperatures of $T_e = (5600\pm130)$\,K and $T_e = (3300\pm120)$\,K, respectively. The measured electron temperatures are found to be systematically low when compared to estimates from radio recombination lines \citep{Paladini2004,Alves2012,Alves2015} and optical line ratios \citep[e.g.,][]{Haffner1999}. Assuming that the true electron temperature of the warm interstellar medium is within the range $6000 < T_e < 8000$\,K we estimate that the degree of H$\alpha$ scattered light \citep{Wood1999,Witt2010} within the H$\alpha$ template is either $0.04 < f_{\mathrm{sc}} < 0.23$ (no dust absorption) or $0.38 < f_{\mathrm{sc}} < 0.64$ ($f_D = 0.33$).

Finally, we have demonstrated that the \mbox{C-BASS} data are an invaluable tool for studying large-scale structures in the ISM. In the future we intend to extend this analysis by looking in more detail at the diffuse ionized gas component of the ISM and placing more rigorous constraints on H$\alpha$ scattering and H\textsc{ii} electron temperatures within the ISM. Further, we wish to extend the technique to include polarization templates of Galactic emission, with the intention of testing magnetic dust emission models of AME and placing more stringent limits on the polarization fraction of AME.

\section*{Acknowledgements}

We thank the referee, Unni Fuskeland, for their useful comments and suggestions that have helped to greatly improve the quality of this paper. The \mbox{C-BASS} project (\url{http://cbass.web.ox.ac.uk}) is a collaboration between Oxford and Manchester Universities in the UK, the California Institute of Technology in the U.S.A., Rhodes University, UKZN and the South African Radio Observatory in South Africa, and the King Abdulaziz City for Science and Technology (KACST) in Saudi Arabia. It has been supported by the NSF awards AST-0607857, AST-1010024, AST-1212217, and AST-1616227, and NASA award NNX15AF06G, the University of Oxford, the Royal Society, STFC, and the other participating institutions. This research was also supported by the South African Radio Astronomy Observatory, which is a facility of the National Research Foundation, an agency of the Department of Science and Technology. We would like to thank Russ Keeney for technical help at OVRO. SEH, CD and JPL acknowledge support from the STFC Consolidated Grant (ST/P000649/1). CD thanks the California Institute of Technology for their hospitality and hosting during several extended visits. We make use of the HEALPix package \citep{gorski2005},  IDL astronomy library \citep{astroIDL1993} and Python astropy \citep{astropy2013,astropy2018}, matplotlib \citep{Hunter:2007}, numpy \citep{2020NumPy-Array}, healpy \citep{gorski2005,Zonca2019}, emcee \citep{Foreman-Mackey2013}, and scipy \citep{2020SciPy-NMeth} packages. This research has made use of the NASA/IPAC Extragalactic Database (NED) which is operated by the Jet Propulsion Laboratory, California Institute of Technology, under contract with the National Aeronautics and Space Administration.

\section*{Data Availability}

The northern \mbox{C-BASS} intensity and polarization data are not currently available, but will be released following the publication of the upcoming survey paper \mbox{(Taylor et al. in prep.)}. Full tables of template fitting coefficients will be made available online upon request.

\bibliography{templatefitting,planck}

\appendix

\vspace{-0.75cm}
\section{Template Fitting Spectral Index Biases}\label{sec:sync_bias}

There are two biases to consider when using \autoref{eqn:sync3} to derive spectral indices. The first is associated with the transformation of the template fitted coefficients probability distribution when transformed into a spectral index. We find that this transformation tends to bias the derived spectral index toward \textit{flatter} indices on average (but also has a long tail towards much steeper indices), which becomes significant when the template coefficients have signal-to-noise ratios less than 5. The second bias is associated with the purity of the synchrotron templates used, for \mbox{C-BASS} we must subtract both the free-free component and the background point source population to avoid biasing our spectral index estimates \textit{flatter}. As such we only use the free-free and source subtracted \mbox{C-BASS} 4.76\,GHz data in this analysis. We have quantified these biases using simulations. 

\begin{figure}
    \centering
    \includegraphics[width=0.4\textwidth]{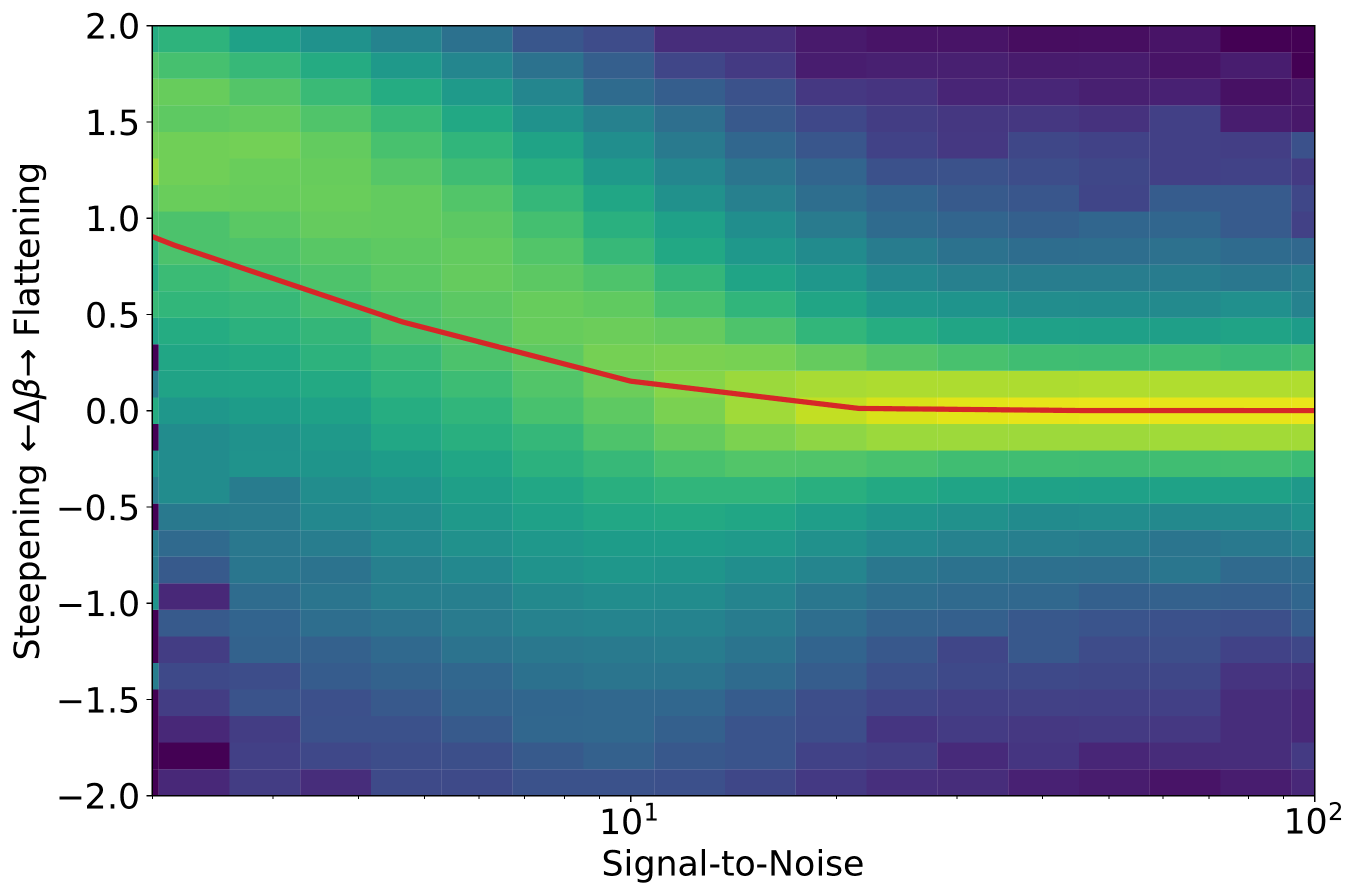}
    \caption{Probability distributions of the change in the measured spectral index for a given template coefficient signal-to-noise ratio. The \textit{red line} indicates the average.}
    \label{fig:bias1}
\end{figure}

When deriving the synchrotron spectral index through template fitting there are two potential forms of bias that must be considered. The first form of bias comes from the transformation of the gaussian probability distribution associated with the fitted template coefficient ($a$) into the non-gaussian probability distribution of the associated synchrotron spectral index through \autoref{eqn:sync3}. In \autoref{fig:bias1} we show how, for a given signal-to-noise on a synchrotron template coefficient, the associated probability distribution and mean change in the measured spectral index. We find that, on average, as the signal-to-noise ratios approach 10 or less, the average spectral index measured flattens, however we also find that the probability distribution skews strongly towards much \textit{steeper} spectral indices. This is a well known effect, and attempts to correct for this bias have been implemented in, for example, the \texttt{COMMANDER} analysis \citep{eriksen2008}. In future analyses we plan to include a similar correction. 

\begin{figure}
    \centering
    \includegraphics[width=0.4\textwidth]{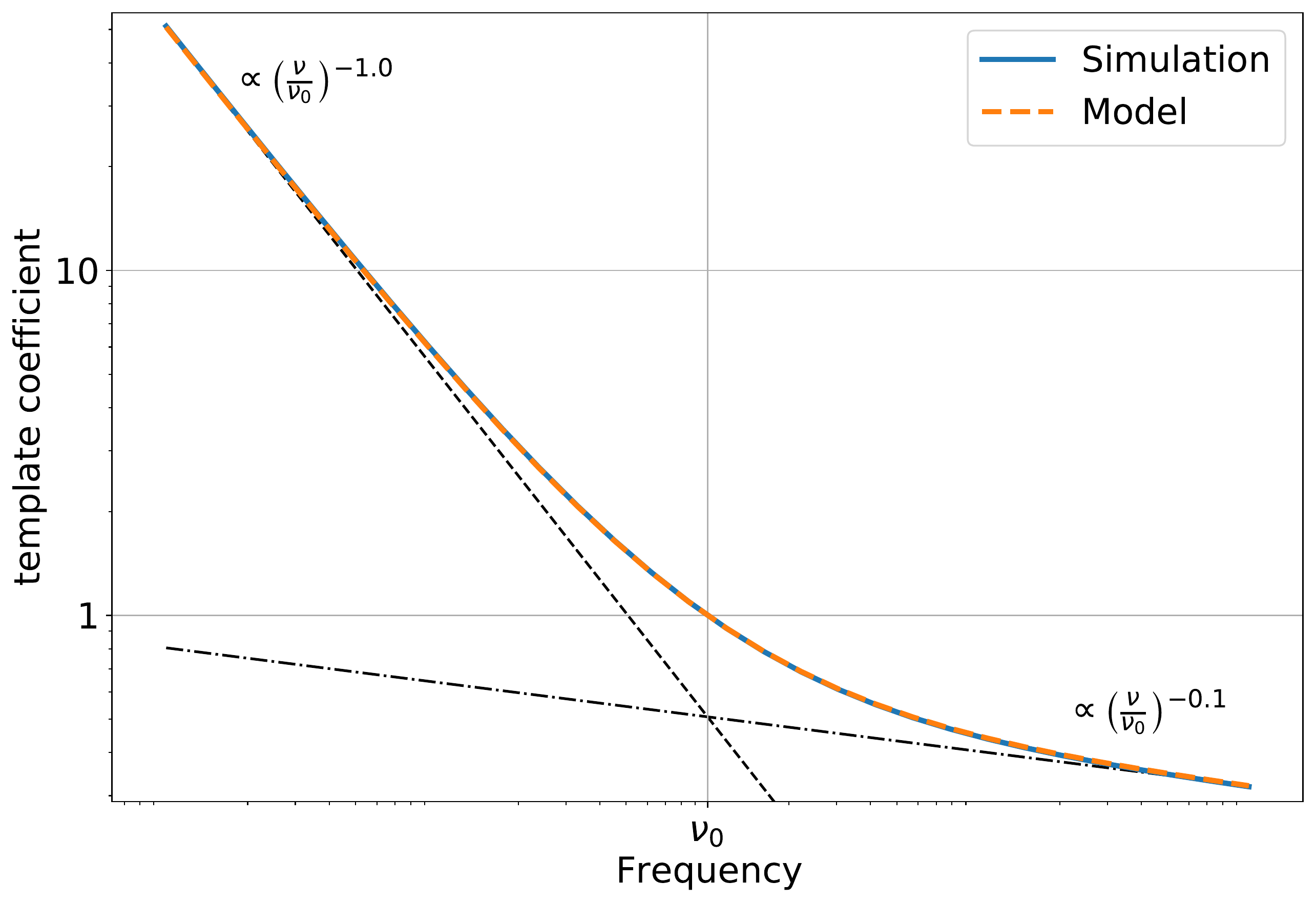}
    \caption{Measured template coefficients when fitting a template that contains two components that correlated with the dataset being fitted. The toy model has a steep spectrum component at low frequencies, and a flat spectrum component at high frequencies.}
    \label{fig:bias2}
\end{figure}

The second form of bias to consider when deriving spectral indices is related to \textit{purity} of the template, i.e. how well does the template represent just a single emission component. In the case of the \mbox{C-BASS} data the map contains a mixture of synchrotron, free-free, and background radio point sources. In the following toy model we will assume we have two components in the template map that are both present in the data map being fitted to. We can define the template as
\begin{equation}
  \mathbf{m} = \mathbf{A}_0 + \mathbf{B}_0,
\end{equation}
where $\mathbf{A}_0$ and $\mathbf{B}_0$ are two uncorrelated brightness distributions. We then fit the template $\mathbf{m}$ to a dataset at a different frequency that can be described by
\begin{equation}
  \mathbf{z} = \mathbf{A}_0\left(\frac{\nu}{\nu_0}\right)^{\beta} + \mathbf{B}_0\left(\frac{\nu}{\nu_0}\right)^{\gamma},
\end{equation}
where $\beta$ and $\gamma$ are the spectral indices of the two components, and $\nu_0$ is an arbitrary reference frequency. We find that, using the template fitting equations from \autoref{sec:templatefitting}, the template coefficient associated with $\mathbf{m}$ is
\begin{equation}\label{eqn:sync_sim3}
  \hat{a}_\mathbf{m} = \frac{1}{\sigma_{A_0}^2 + \sigma^2_{B_{0}}}\left[\sigma_{A_0}^2\left(\frac{\nu}{\nu_0}\right)^{\beta} +\sigma_{B_0}^2\left(\frac{\nu}{\nu_0}\right)^{\gamma} \right],
\end{equation}
where we can see the associated template coefficient is the sum of two power-laws. Therefore assuming that $\beta < \gamma$, we will expect that at frequencies $\nu \ll \nu_0$ the power-law has a spectral index of $\beta$ and at high frequencies ($\nu \gg \nu_0$) it will follow $\gamma$. 

A toy model of this effect can be seen in \autoref{fig:bias2}, which compares simulations (solid \textit{blue} line) with the model described by \autoref{eqn:sync_sim3} (dashed \textit{orange} line). The plot shows the template coefficient that would be measured between the template frequency and the data frequency. We can see at high frequencies the flat spectrum component is the dominant contribution and the spectral index that would be measured converges to $\gamma = -0.1$, while at low frequencies it would be dominated by the steep spectrum component, i.e. $\beta = -1$. This model demonstrates the importance of ensuring that the templates used only contain a single emission component otherwise it can severely bias the resulting interpretation of the fitted coefficients.

\vspace{-0.75cm}
\section{Template Fitting Coefficients at 22.8GHz}

\begin{table*}
\centering
\caption{Coefficients measured for each region at 22.8\,GHz. Coefficients use the \mbox{C-BASS} 4.76\,GHz synchrotron template, $\tau_{353}$ dust template, and the $f_D=0.33$ H$\alpha$ template. Synchrotron spectral indices are given for both 4.76\,GHz and 408\,MHz templates. Each row gives the approximate central Galactic longitude and latitude.}
\label{tab:all_coeff1}
\begin{tabular}{c|c|c|c|c|c|c|c|c}
\hline
\hline
Region & Sync. & Dust & H$\alpha$ & $\chi_r^2$ & $\beta_{s,4.76}$ & $\beta_{s,408}$ & $\ell$ & $b$\\ 
 & mK/K & K/$\tau_{353}$ & $\mu$K/R &  &  &  & deg. & deg.\\ \hline
001 & $8.4 \pm 0.4$ & $9.7 \pm 0.4$ & $4.7 \pm 1.9$ & 1.8 & $-3.05 \pm 0.03$ & $-2.89 \pm 0.01$ & $130$ & $ 20$\\ 
002 & $5.1 \pm 0.9$ & $11.0 \pm 0.5$ & $-27.5 \pm 6.2$ & 2.7 & $-3.37 \pm 0.12$ & $-2.96 \pm 0.04$ & $130$ & $ 30$\\ 
003 & $4.7 \pm 1.2$ & $7.8 \pm 0.5$ & $4.8 \pm 4.2$ & 1.4 & $-3.41 \pm 0.17$ & $-3.13 \pm 0.05$ & $120$ & $ 40$\\ 
004 & $7.6 \pm 0.8$ & $7.0 \pm 0.7$ & $31.6 \pm 3.7$ & 9.2 & $-3.11 \pm 0.07$ & $-2.91 \pm 0.03$ & $110$ & $ 20$\\ 
005 & $8.1 \pm 1.9$ & $11.0 \pm 0.7$ & $8.6 \pm 1.0$ & 4.7 & $-3.07 \pm 0.15$ & $-3.15 \pm 0.11$ & $140$ & $ 20$\\ 
006 & $9.9 \pm 0.7$ & $9.0 \pm 1.0$ & $3.6 \pm 2.1$ & 1.5 & $-2.95 \pm 0.04$ & $-2.94 \pm 0.02$ & $150$ & $ 30$\\ 
007 & $11.8 \pm 1.7$ & $15.7 \pm 0.9$ & $-18.5 \pm 3.8$ & 1.6 & $-2.83 \pm 0.09$ & $-3.0 \pm 0.03$ & $140$ & $ 40$\\ 
008 & $5.8 \pm 1.1$ & $3.1 \pm 2.2$ & $10.1 \pm 5.1$ & 0.8 & $-3.29 \pm 0.12$ & $-3.39 \pm 0.16$ & $120$ & $ 50$\\ 
009 & $11.3 \pm 1.4$ & $14.8 \pm 2.1$ & $-4.3 \pm 3.3$ & 1.5 & $-2.86 \pm 0.08$ & $-3.02 \pm 0.03$ & $100$ & $ 40$\\ 
010 & $9.9 \pm 1.3$ & $8.1 \pm 0.8$ & $18.1 \pm 3.5$ & 6.2 & $-2.94 \pm 0.08$ & $-2.99 \pm 0.03$ & $100$ & $ 20$\\ 
011 & $5.3 \pm 0.8$ & $10.8 \pm 0.6$ & $12.4 \pm 0.8$ & 5.4 & $-3.35 \pm 0.1$ & $-3.04 \pm 0.05$ & $130$ & $-10$\\ 
012 & $5.1 \pm 1.2$ & $10.9 \pm 0.6$ & $10.8 \pm 1.8$ & 4.1 & $-3.38 \pm 0.15$ & $-3.1 \pm 0.1$ & $150$ & $ 10$\\ 
013 & $6.2 \pm 1.3$ & $11.8 \pm 0.8$ & $7.5 \pm 4.2$ & 1.6 & $-3.24 \pm 0.13$ & $-3.14 \pm 0.04$ & $160$ & $ 20$\\ 
014 & $13.6 \pm 0.9$ & $11.9 \pm 0.9$ & $4.0 \pm 5.0$ & 1.0 & $-2.74 \pm 0.04$ & $-2.97 \pm 0.02$ & $160$ & $ 40$\\ 
015 & $16.0 \pm 2.3$ & $36.3 \pm 4.1$ & $-14.6 \pm 2.8$ & 1.3 & $-2.64 \pm 0.09$ & $-2.95 \pm 0.02$ & $150$ & $ 60$\\ 
016 & $2.2 \pm 1.0$ & $23.8 \pm 7.0$ & $-27.7 \pm 8.7$ & 1.1 & $-3.9 \pm 0.3$ & $-3.4 \pm 0.23$ & $120$ & $ 60$\\ 
017 & $9.6 \pm 1.4$ & $5.7 \pm 4.4$ & $-9.2 \pm 4.3$ & 0.8 & $-2.97 \pm 0.09$ & $-3.07 \pm 0.04$ & $ 90$ & $ 50$\\ 
018 & $8.7 \pm 1.1$ & $21.0 \pm 1.7$ & $-9.2 \pm 4.8$ & 1.4 & $-3.03 \pm 0.08$ & $-3.06 \pm 0.03$ & $ 80$ & $ 40$\\ 
019 & $6.5 \pm 0.4$ & $14.3 \pm 0.9$ & $4.2 \pm 0.5$ & 1.3 & $-3.22 \pm 0.04$ & $-2.99 \pm 0.02$ & $ 80$ & $ 20$\\ 
020 & $2.6 \pm 1.3$ & $13.3 \pm 1.1$ & $6.7 \pm 1.1$ & 4.6 & $-3.79 \pm 0.32$ & $-3.4 \pm 0.32$ & $100$ & $-10$\\ 
021 & $7.6 \pm 0.9$ & $12.6 \pm 0.9$ & $8.4 \pm 0.6$ & 1.4 & $-3.12 \pm 0.08$ & $-2.96 \pm 0.03$ & $120$ & $-20$\\ 
022 & $7.3 \pm 0.4$ & $13.7 \pm 0.9$ & $7.2 \pm 0.8$ & 1.6 & $-3.14 \pm 0.04$ & $-2.96 \pm 0.02$ & $140$ & $-20$\\ 
023 & $-7.1 \pm 3.1$ & $10.0 \pm 1.4$ & $15.7 \pm 1.5$ & 77.9 & -- & -- & $160$ & $-10$\\ 
024 & $1.2 \pm 1.7$ & $9.3 \pm 0.6$ & $16.1 \pm 2.0$ & 15.3 & $-4.28 \pm 0.9$ & -- & $170$ & $ 10$\\ 
025 & $0.8 \pm 1.4$ & $13.1 \pm 0.7$ & $2.5 \pm 2.5$ & 2.9 & $-4.59 \pm 1.22$ & -- & $170$ & $ 20$\\ 
026 & $5.1 \pm 0.8$ & $16.6 \pm 1.7$ & $5.0 \pm 2.0$ & 1.5 & $-3.36 \pm 0.1$ & $-3.15 \pm 0.03$ & $180$ & $ 30$\\ 
027 & $11.8 \pm 1.4$ & $25.5 \pm 5.3$ & $-9.8 \pm 10.4$ & 2.7 & $-2.83 \pm 0.08$ & $-2.92 \pm 0.03$ & $180$ & $ 50$\\ 
028 & $9.3 \pm 1.1$ & $24.3 \pm 5.8$ & $4.4 \pm 6.3$ & 1.4 & $-2.98 \pm 0.08$ & $-3.06 \pm 0.04$ & $170$ & $ 70$\\ 
029 & $7.4 \pm 0.8$ & $24.1 \pm 3.9$ & $-13.0 \pm 5.4$ & 0.7 & $-3.13 \pm 0.07$ & $-3.28 \pm 0.09$ & $130$ & $ 70$\\ 
030 & $11.5 \pm 1.0$ & $15.5 \pm 4.0$ & $-9.3 \pm 6.0$ & 0.8 & $-2.85 \pm 0.05$ & $-3.02 \pm 0.03$ & $ 80$ & $ 70$\\ 
031 & $7.0 \pm 1.0$ & $15.5 \pm 2.6$ & $-8.5 \pm 5.2$ & 0.8 & $-3.17 \pm 0.09$ & $-3.04 \pm 0.02$ & $ 70$ & $ 50$\\ 
032 & $10.3 \pm 2.0$ & $12.8 \pm 3.5$ & $-13.8 \pm 6.1$ & 4.9 & $-2.92 \pm 0.12$ & $-3.33 \pm 0.16$ & $ 70$ & $ 30$\\ 
033 & $6.3 \pm 0.5$ & $15.0 \pm 0.6$ & $4.5 \pm 1.0$ & 1.4 & $-3.24 \pm 0.05$ & $-3.12 \pm 0.03$ & $ 70$ & $ 20$\\ 
034 & $6.5 \pm 1.0$ & $13.4 \pm 0.9$ & $5.1 \pm 0.8$ & 5.7 & $-3.22 \pm 0.09$ & $-3.0 \pm 0.04$ & $ 90$ & $-10$\\ 
035 & $9.2 \pm 1.2$ & $12.9 \pm 0.7$ & $8.1 \pm 0.7$ & 3.4 & $-2.99 \pm 0.08$ & $-3.07 \pm 0.06$ & $110$ & $-20$\\ 
036 & $6.0 \pm 1.0$ & $14.5 \pm 0.8$ & $-1.9 \pm 4.7$ & 1.0 & $-3.26 \pm 0.11$ & $-2.99 \pm 0.04$ & $110$ & $-30$\\ 
037 & $9.7 \pm 1.1$ & $12.8 \pm 0.8$ & $-2.9 \pm 2.5$ & 1.7 & $-2.96 \pm 0.08$ & $-3.01 \pm 0.07$ & $130$ & $-30$\\ 
038 & $2.6 \pm 1.7$ & $10.3 \pm 1.2$ & $18.7 \pm 3.4$ & 14.7 & $-3.81 \pm 0.42$ & -- & $150$ & $-30$\\ 
039 & $17.7 \pm 1.6$ & $5.2 \pm 0.4$ & $2.8 \pm 1.3$ & 30.4 & $-2.57 \pm 0.06$ & $-2.71 \pm 0.02$ & $170$ & $-10$\\ 
040 & $16.2 \pm 4.0$ & $10.6 \pm 0.4$ & $1.2 \pm 1.8$ & 15.4 & $-2.63 \pm 0.16$ & $-2.93 \pm 0.09$ & $180$ & $ 10$\\ 
041 & $6.6 \pm 0.8$ & $8.6 \pm 1.2$ & $4.7 \pm 1.6$ & 1.3 & $-3.21 \pm 0.08$ & $-3.09 \pm 0.04$ & $190$ & $ 20$\\ 
042 & $3.5 \pm 2.0$ & $12.9 \pm 2.1$ & $10.5 \pm 5.0$ & 1.7 & $-3.61 \pm 0.37$ & $-3.49 \pm 0.3$ & $200$ & $ 40$\\ 
043 & $6.0 \pm 1.0$ & $15.9 \pm 2.6$ & $-1.1 \pm 5.1$ & 1.0 & $-3.27 \pm 0.11$ & $-3.07 \pm 0.05$ & $200$ & $ 60$\\ 
044 & $7.4 \pm 1.5$ & $31.8 \pm 5.9$ & $-7.7 \pm 7.5$ & 1.2 & $-3.13 \pm 0.13$ & $-3.24 \pm 0.09$ & $200$ & $ 80$\\ 
045 & $4.5 \pm 1.1$ & $43.5 \pm 7.7$ & $-12.6 \pm 13.6$ & 2.0 & $-3.45 \pm 0.15$ & $-3.23 \pm 0.09$ & $ 50$ & $ 80$\\ 
046 & $9.8 \pm 0.4$ & $19.5 \pm 1.3$ & $-18.0 \pm 4.5$ & 0.7 & $-2.95 \pm 0.03$ & $-3.02 \pm 0.01$ & $ 50$ & $ 60$\\ 
047 & $8.2 \pm 1.1$ & $19.3 \pm 1.8$ & $1.8 \pm 5.1$ & 3.0 & $-3.07 \pm 0.09$ & $-3.15 \pm 0.04$ & $ 50$ & $ 40$\\ 
048 & $6.6 \pm 0.6$ & $15.7 \pm 0.8$ & $-0.2 \pm 1.2$ & 1.7 & $-3.21 \pm 0.06$ & $-3.07 \pm 0.02$ & $ 60$ & $ 20$\\ 
049 & $7.1 \pm 0.6$ & $11.1 \pm 0.5$ & $9.7 \pm 0.5$ & 3.2 & $-3.16 \pm 0.05$ & $-2.99 \pm 0.03$ & $ 60$ & $ 10$\\ 
050 & $5.6 \pm 1.4$ & $7.9 \pm 0.7$ & $7.7 \pm 0.7$ & 3.7 & $-3.31 \pm 0.16$ & $-3.43 \pm 0.19$ & $ 70$ & $-10$\\ 
051 & $5.8 \pm 0.7$ & $15.7 \pm 1.9$ & $6.6 \pm 0.5$ & 2.3 & $-3.29 \pm 0.08$ & $-3.09 \pm 0.04$ & $ 90$ & $-20$\\ 
052 & $3.9 \pm 1.3$ & $14.3 \pm 0.9$ & $0.4 \pm 5.1$ & 1.3 & $-3.54 \pm 0.22$ & $-3.07 \pm 0.07$ & $120$ & $-40$\\ 
053 & $8.9 \pm 0.9$ & $12.0 \pm 0.9$ & $-0.3 \pm 4.6$ & 2.7 & $-3.02 \pm 0.06$ & $-2.85 \pm 0.02$ & $150$ & $-40$\\ 
054 & $14.3 \pm 1.3$ & $14.1 \pm 0.7$ & $5.1 \pm 1.0$ & 13.4 & $-2.71 \pm 0.06$ & $-2.8 \pm 0.02$ & $170$ & $-30$\\ \hline
\end{tabular}
\end{table*}

\begin{table*} 
\centering
\caption{\autoref{tab:all_coeff1} continued.}
\label{tab:all_coeff2}
\begin{tabular}{c|c|c|c|c|c|c|c|c}
\hline
\hline
Region & Sync. & Dust & H$\alpha$ & $\chi_r^2$ & $\beta_{s,4.76}$ & $\beta_{s,408}$ & $\ell$ & $b$\\ 
 & mK/K & K/$\tau_{353}$ & $\mu$K/R &  &  &  & deg. & deg.\\ \hline
055 & $-7.6 \pm 1.5$ & $11.8 \pm 0.8$ & $8.7 \pm 1.2$ & 36.8 & -- & -- & $180$ & $-10$\\ 
056 & $8.9 \pm 1.1$ & $9.9 \pm 0.4$ & $6.6 \pm 0.7$ & 4.8 & $-3.02 \pm 0.08$ & $-2.99 \pm 0.05$ & $200$ & $ 10$\\ 
057 & $7.1 \pm 0.6$ & $9.3 \pm 2.1$ & $5.3 \pm 1.3$ & 1.8 & $-3.16 \pm 0.05$ & $-3.04 \pm 0.02$ & $200$ & $ 30$\\ 
058 & $9.3 \pm 0.8$ & $10.6 \pm 1.6$ & $8.3 \pm 2.7$ & 0.8 & $-2.98 \pm 0.06$ & $-3.03 \pm 0.03$ & $210$ & $ 50$\\ 
059 & $3.8 \pm 1.1$ & $10.6 \pm 1.5$ & $-0.9 \pm 4.3$ & 1.1 & $-3.57 \pm 0.19$ & $-3.12 \pm 0.06$ & $230$ & $ 70$\\ 
060 & $8.8 \pm 1.2$ & $11.0 \pm 1.7$ & $-2.1 \pm 13.7$ & 3.0 & $-3.02 \pm 0.09$ & $-3.06 \pm 0.04$ & $-70$ & $ 80$\\ 
061 & $8.3 \pm 0.3$ & $23.3 \pm 2.8$ & $8.6 \pm 6.4$ & 1.4 & $-3.06 \pm 0.03$ & $-3.02 \pm 0.01$ & $ 10$ & $ 70$\\ 
062 & $7.0 \pm 0.3$ & $15.1 \pm 1.5$ & $2.1 \pm 4.9$ & 1.3 & $-3.17 \pm 0.03$ & $-3.01 \pm 0.01$ & $ 30$ & $ 50$\\ 
063 & $5.8 \pm 0.3$ & $9.2 \pm 1.0$ & $28.3 \pm 4.2$ & 1.1 & $-3.28 \pm 0.04$ & $-3.07 \pm 0.01$ & $ 40$ & $ 30$\\ 
064 & $6.8 \pm 0.9$ & $9.2 \pm 0.7$ & $13.5 \pm 1.1$ & 5.1 & $-3.19 \pm 0.09$ & $-2.99 \pm 0.02$ & $ 60$ & $-10$\\ 
065 & $6.1 \pm 0.6$ & $14.3 \pm 0.8$ & $5.0 \pm 1.1$ & 1.8 & $-3.25 \pm 0.06$ & $-3.22 \pm 0.06$ & $ 70$ & $-20$\\ 
066 & $12.9 \pm 2.2$ & $14.7 \pm 0.7$ & $2.7 \pm 8.7$ & 9.3 & $-2.78 \pm 0.11$ & $-2.87 \pm 0.05$ & $ 90$ & $-40$\\ 
067 & $8.4 \pm 1.0$ & $14.1 \pm 0.7$ & $11.0 \pm 4.5$ & 2.1 & $-3.05 \pm 0.07$ & $-2.97 \pm 0.03$ & $100$ & $-50$\\ 
068 & $6.2 \pm 0.7$ & $15.8 \pm 1.1$ & $-21.7 \pm 7.1$ & 1.4 & $-3.24 \pm 0.07$ & $-2.99 \pm 0.03$ & $140$ & $-50$\\ 
069 & $9.8 \pm 1.8$ & $15.6 \pm 0.6$ & $1.6 \pm 1.3$ & 3.9 & $-2.95 \pm 0.12$ & $-3.3 \pm 0.1$ & $170$ & $-40$\\ 
070 & $16.0 \pm 1.3$ & $13.0 \pm 0.8$ & $4.7 \pm 0.7$ & 16.5 & $-2.64 \pm 0.05$ & $-2.85 \pm 0.03$ & $190$ & $-30$\\ 
071 & $6.6 \pm 3.3$ & $5.7 \pm 1.1$ & $9.9 \pm 0.7$ & 82.8 & $-3.21 \pm 0.32$ & $-3.6 \pm 1.09$ & $200$ & $-10$\\ 
072 & $4.0 \pm 0.9$ & $9.0 \pm 0.7$ & $11.9 \pm 1.1$ & 3.6 & $-3.52 \pm 0.14$ & $-3.12 \pm 0.05$ & $210$ & $ 10$\\ 
073 & $5.0 \pm 1.2$ & $12.4 \pm 1.7$ & $9.0 \pm 1.9$ & 1.9 & $-3.38 \pm 0.15$ & $-3.01 \pm 0.05$ & $220$ & $ 30$\\ 
074 & $5.6 \pm 1.8$ & $8.4 \pm 1.9$ & $2.0 \pm 3.0$ & 1.8 & $-3.3 \pm 0.21$ & $-3.31 \pm 0.12$ & $230$ & $ 50$\\ 
075 & $9.8 \pm 3.4$ & $8.5 \pm 1.6$ & $-23.2 \pm 8.4$ & 8.8 & $-2.95 \pm 0.22$ & -- & $270$ & $ 70$\\ 
076 & $7.3 \pm 0.6$ & $-2.6 \pm 2.2$ & $-5.2 \pm 6.6$ & 0.7 & $-3.14 \pm 0.05$ & $-3.13 \pm 0.02$ & $330$ & $ 70$\\ 
077 & $6.0 \pm 0.4$ & $9.3 \pm 2.2$ & $20.0 \pm 4.9$ & 1.4 & $-3.27 \pm 0.05$ & $-3.12 \pm 0.02$ & $ 10$ & $ 60$\\ 
078 & $5.2 \pm 0.4$ & $12.8 \pm 1.3$ & $-0.6 \pm 3.9$ & 1.1 & $-3.36 \pm 0.05$ & $-3.17 \pm 0.02$ & $ 20$ & $ 40$\\ 
079 & $5.8 \pm 0.4$ & $12.0 \pm 0.9$ & $14.0 \pm 2.4$ & 1.8 & $-3.28 \pm 0.04$ & $-3.08 \pm 0.03$ & $ 60$ & $-20$\\ 
080 & $7.4 \pm 0.9$ & $11.4 \pm 1.0$ & $14.7 \pm 3.3$ & 2.5 & $-3.13 \pm 0.07$ & $-3.1 \pm 0.03$ & $ 70$ & $-40$\\ 
081 & $10.2 \pm 0.9$ & $13.9 \pm 1.5$ & $32.4 \pm 9.1$ & 1.8 & $-2.93 \pm 0.05$ & $-3.03 \pm 0.04$ & $120$ & $-60$\\ 
082 & $7.7 \pm 0.9$ & $15.7 \pm 1.0$ & $-15.7 \pm 4.4$ & 1.0 & $-3.11 \pm 0.07$ & $-3.04 \pm 0.03$ & $160$ & $-60$\\ 
083 & $8.0 \pm 1.3$ & $15.6 \pm 0.7$ & $4.5 \pm 0.4$ & 7.9 & $-3.08 \pm 0.11$ & $-3.42 \pm 0.33$ & $190$ & $-40$\\ 
084 & $6.6 \pm 2.7$ & $22.0 \pm 3.0$ & $5.4 \pm 0.8$ & 21.7 & $-3.2 \pm 0.26$ & $-3.04 \pm 0.1$ & $200$ & $-20$\\ 
085 & $11.2 \pm 1.7$ & $7.5 \pm 0.5$ & $4.5 \pm 0.8$ & 11.3 & $-2.87 \pm 0.1$ & $-3.12 \pm 0.13$ & $210$ & $ -0$\\ 
086 & $5.7 \pm 1.0$ & $13.3 \pm 1.2$ & $-1.6 \pm 1.9$ & 1.5 & $-3.3 \pm 0.11$ & $-3.39 \pm 0.16$ & $220$ & $ 20$\\ 
087 & $6.2 \pm 1.5$ & $12.7 \pm 1.1$ & $4.9 \pm 2.7$ & 1.4 & $-3.25 \pm 0.16$ & -- & $240$ & $ 40$\\ 
088 & $9.7 \pm 2.1$ & $9.2 \pm 1.2$ & $7.2 \pm 2.4$ & 2.6 & $-2.96 \pm 0.14$ & $-3.05 \pm 0.07$ & $260$ & $ 50$\\ 
089 & $8.9 \pm 2.3$ & $31.7 \pm 22.8$ & $32.0 \pm 14.9$ & 47.2 & $-3.01 \pm 0.17$ & $-3.08 \pm 0.08$ & $300$ & $ 60$\\ 
090 & $6.9 \pm 0.7$ & $11.9 \pm 1.0$ & $4.7 \pm 1.3$ & 0.9 & $-3.18 \pm 0.06$ & $-3.16 \pm 0.04$ & $340$ & $ 60$\\ 
091 & $4.1 \pm 1.0$ & $8.6 \pm 1.8$ & $56.2 \pm 12.7$ & 3.8 & $-3.51 \pm 0.16$ & $-3.15 \pm 0.06$ & $ 10$ & $ 40$\\ 
092 & $5.0 \pm 0.4$ & $14.6 \pm 0.6$ & $2.5 \pm 1.5$ & 2.4 & $-3.38 \pm 0.05$ & $-3.15 \pm 0.03$ & $ 40$ & $-20$\\ 
093 & $5.1 \pm 0.5$ & $15.7 \pm 0.9$ & $4.2 \pm 5.2$ & 6.6 & $-3.37 \pm 0.06$ & $-3.08 \pm 0.03$ & $ 60$ & $-40$\\ 
094 & $7.6 \pm 1.0$ & $14.4 \pm 0.9$ & $7.2 \pm 3.8$ & 1.1 & $-3.11 \pm 0.08$ & $-3.1 \pm 0.04$ & $ 80$ & $-50$\\ 
095 & $7.9 \pm 0.8$ & $8.3 \pm 2.8$ & $-1.7 \pm 8.7$ & 1.8 & $-3.09 \pm 0.06$ & $-3.01 \pm 0.03$ & $ 90$ & $-70$\\ 
096 & $3.1 \pm 1.1$ & $23.5 \pm 1.1$ & $14.3 \pm 5.0$ & 1.3 & $-3.69 \pm 0.22$ & $-2.97 \pm 0.03$ & $150$ & $-70$\\ 
097 & $4.9 \pm 0.5$ & $14.6 \pm 0.8$ & $5.2 \pm 0.3$ & 0.9 & $-3.4 \pm 0.07$ & $-2.95 \pm 0.03$ & $190$ & $-50$\\ 
098 & $1.3 \pm 2.2$ & $22.0 \pm 1.4$ & $5.6 \pm 0.7$ & 3.6 & $-4.26 \pm 1.13$ & -- & $200$ & $-30$\\ 
099 & $12.1 \pm 1.9$ & $7.0 \pm 0.5$ & $4.6 \pm 0.4$ & 22.1 & $-2.82 \pm 0.1$ & $-2.94 \pm 0.12$ & $220$ & $-10$\\ 
100 & $3.4 \pm 1.6$ & $7.9 \pm 0.4$ & $8.5 \pm 0.9$ & 13.0 & $-3.62 \pm 0.29$ & $-3.15 \pm 0.11$ & $220$ & $ 10$\\ 
101 & $2.7 \pm 1.4$ & $18.2 \pm 1.2$ & $8.5 \pm 2.3$ & 1.1 & $-3.78 \pm 0.33$ & $-3.19 \pm 0.06$ & $240$ & $ 20$\\ 
102 & $7.9 \pm 0.7$ & $10.1 \pm 1.0$ & $-1.8 \pm 2.8$ & 0.7 & $-3.09 \pm 0.06$ & $-3.26 \pm 0.12$ & $250$ & $ 40$\\ 
103 & $6.7 \pm 0.5$ & $21.7 \pm 1.9$ & $-28.5 \pm 9.4$ & 1.5 & $-3.2 \pm 0.05$ & $-2.92 \pm 0.02$ & $280$ & $ 50$\\ 
104 & $4.7 \pm 2.3$ & $31.7 \pm 4.0$ & $13.9 \pm 3.1$ & 4.3 & $-3.42 \pm 0.31$ & $-3.06 \pm 0.1$ & $320$ & $ 50$\\ 
105 & $9.6 \pm 1.1$ & $16.9 \pm 1.3$ & $-1.1 \pm 10.4$ & 4.3 & $-2.97 \pm 0.08$ & $-2.87 \pm 0.03$ & $350$ & $ 40$\\ 
106 & $-3.4 \pm 2.3$ & $21.5 \pm 1.7$ & $12.4 \pm 1.5$ & 19.3 & -- & -- & $ 10$ & $ 30$\\ 
107 & $5.5 \pm 1.0$ & $12.3 \pm 0.8$ & $5.5 \pm 3.9$ & 1.5 & $-3.32 \pm 0.11$ & $-3.07 \pm 0.04$ & $ 40$ & $-30$\\ 
108 & $6.1 \pm 2.6$ & $26.1 \pm 6.4$ & $-25.3 \pm 14.9$ & 9.8 & $-3.25 \pm 0.28$ & $-3.21 \pm 0.12$ & $ 60$ & $-50$\\  \hline
\end{tabular}
\end{table*}

The coefficients for all regions at 22.8\,GHz are given in \autoref{tab:all_coeff1} and \autoref{tab:all_coeff2} for the 4.76\,GHz synchrotron template, the $\tau_{353}$ dust template, the H$\alpha$ ($f_D=0.33$) free-free template,  and the reduced $\chi^2_r$ value of each template fit. The tables also include the spectral indices derived between the 408\,MHz or 4.76\,GHz and the 22.8\,GHz data. For some regions no valid spectral index could be derived due to the synchrotron coefficient being negative.

\label{lastpage}

\end{document}